\documentclass[sigconf]{acmart}

\usepackage{acmart-taps}

\usepackage{graphicx}
\usepackage{xspace}
\usepackage{listings}
\usepackage{url}
\usepackage{multirow}
\usepackage{subcaption}

\usepackage{algorithm} 
\usepackage{algorithmic}

\usepackage{url}
\usepackage{hyperref}
\usepackage{hyperxmp}
\usepackage{balance}

\usepackage{cleveref}
\usepackage{booktabs}

\renewcommand{\arraystretch}{1.2}

\AtBeginDocument{%
  \providecommand\BibTeX{{%
    \normalfont B\kern-0.5em{\scshape i\kern-0.25em b}\kern-0.8em\TeX}}}

\copyrightyear{2025}
\acmYear{2025}
\setcopyright{cc}
\setcctype{by}
\acmConference[UIST '25]{The 38th Annual ACM Symposium on User Interface Software and Technology}{September 28-October 1, 2025}{Busan, Republic of Korea}
\acmBooktitle{The 38th Annual ACM Symposium on User Interface Software and Technology (UIST '25), September 28-October 1, 2025, Busan, Republic of Korea}
\acmDOI{10.1145/3746059.3747624}
\acmISBN{979-8-4007-2037-6/2025/09}

\newcommand{\revised}[1]{#1}

\def\systemname {Accessibility Scout\xspace}
\def\eg {\textit{e.g.}\xspace}
\def\ie {\textit{i.e.}\xspace}

\def\etal {\textit{et al.}\xspace}

\begin{document}

\title{Accessibility Scout: Personalized Accessibility Scans \\ of Built Environments}

\author{William Huang}
\orcid{0000-0001-7651-2190}
\affiliation{%
  \institution{University of California, Los Angeles}
  \city{Los Angeles}
  \state{CA}
  \country{USA}
}
\email{william.huang@ucla.edu}

\author{Xia Su}
\orcid{0000-0002-2063-0769}
\affiliation{%
  \institution{University of Washington}
  \city{Seattle}
  \state{WA}
  \country{USA}
}
\email{xiasu@cs.washington.edu}

\author{Jon E. Froehlich}
\orcid{0000-0001-8291-3353}
\affiliation{%
  \institution{University of Washington}
  \city{Seattle}
  \state{WA}
  \country{USA}
}
\email{jonf@cs.uw.edu}

\author{Yang Zhang}
\orcid{0000-0003-2472-6968}
\affiliation{%
  \institution{University of California, Los Angeles}
  \city{Los Angeles}
  \state{CA}
  \country{USA}
}
\email{yangzhang@ucla.edu}

\renewcommand{\shortauthors}{Huang et al.}

\begin{teaserfigure}
  \includegraphics[width=\textwidth]{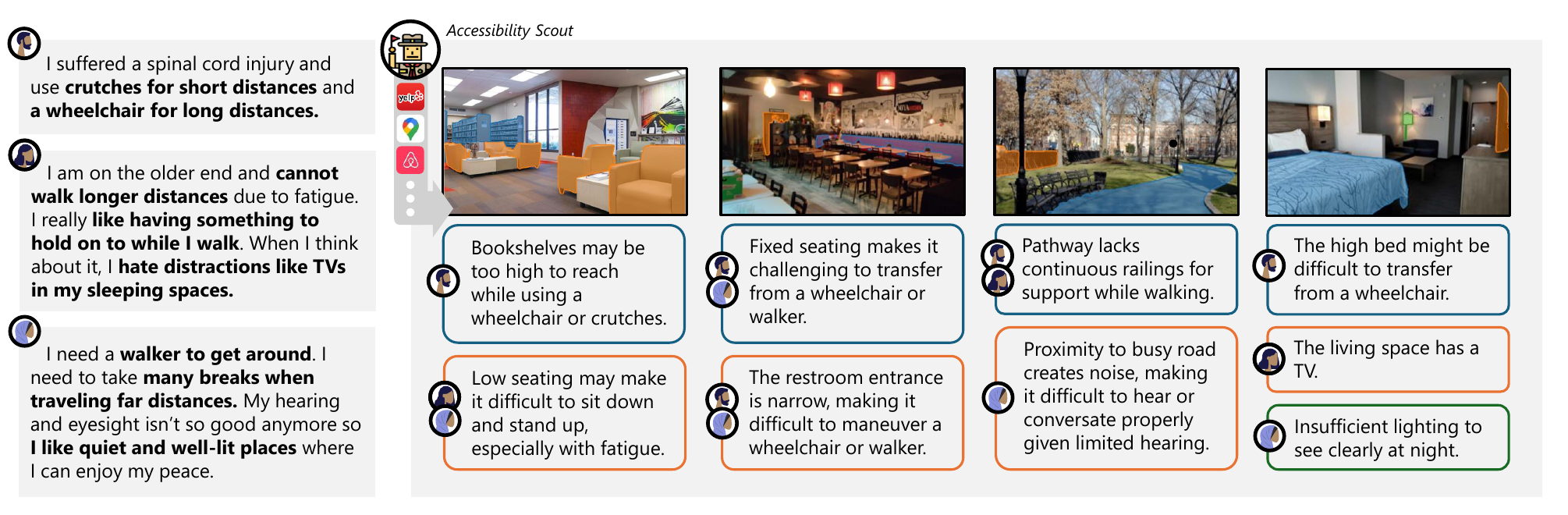}
  \vspace{-0.6cm}
      \caption[]{\systemname is an LLM-based personalized accessibility scanning system for semi-automatically \textit{modeling} a person's accessibility preferences to \textit{identify} and \textit{visualize} accessibility concerns in images of built environments. (Left) \systemname converts plain text descriptions of accessibility and collaborative human-AI accessibility annotations into LLM-interpretable user models. (Right) User models are used to generate personalized accessibility scans for each individual. Images can be sourced from anywhere, including Yelp, Google Maps, AirBnB, Facebook, Booking.com, and more. }
  \Description{ Visualization of system in action. On the left are example user personas for three people: (1) Someone with a spinal cord injury who uses crutches and a wheelchair, (2) someone who is older, cannot walk long distances, likes having something to hold on to, and hates distractions in sleeping spaces, and (3) somone who uses a walker, needs breaks while traveling long distances, likes quiet places, and likes well-lit areas. On the right are identified concerns from the system with different concerns identified for each user. In a library, the system finds that the bookshelves are too high for persona 1 while the seating is too low for personas 2 and 3. In a restaurant, the system finds the fixed seating difficult to transfer and restroom entrance too narrow for personas 1 and 2. In a park, the system finds a lack of supporting rails for persona 1 and 2 and a busy loud street for persona 3. In a hotel room, the system finds the bed too high for persona 1, a distracting TV for persona 2, and insufficient lighting for persona 3.}
  \vspace{0.2cm}
  \label{fig:teaser}
\end{teaserfigure}

\begin{abstract}
Assessing the accessibility of unfamiliar built environments is critical for people with disabilities. However, manual assessments, performed by users or their personal health professionals, are laborious and unscalable, while automatic machine learning methods often neglect an individual user's unique needs. Recent advances in Large Language Models (LLMs) enable novel approaches to this problem,  balancing personalization with scalability to enable more adaptive and context-aware assessments of accessibility. We present \systemname, an LLM-based accessibility scanning system that identifies accessibility concerns from photos of built environments. With use, \systemname becomes an increasingly capable "accessibility scout", tailoring accessibility scans to an individual's mobility level, preferences, and specific environmental interests through collaborative Human-AI assessments. We present findings from three studies: a formative study with six participants to inform the design of \systemname, a technical evaluation of 500 images of built environments, and a user study with 10 participants of varying mobility. Results from our technical evaluation and user study show that \systemname can generate personalized accessibility scans that extend beyond traditional ADA considerations. Finally, we conclude with a discussion on the implications of our work and future steps for building more scalable and personalized accessibility assessments of the physical world.

\end{abstract}



\maketitle


\section{Introduction}

Safe and accessible spaces are crucial for human well-being  \cite{choAccessibleHomeEnvironments2016, liAreAccessibilityFacility2023, gaoRevisitingImpactPublic2024, kapsalisDisabledbydesignEffectsInaccessible2024} and quality of life \cite{potPerceivedAccessibilityWhat2021}. However, these spaces are not guaranteed, especially for people with limited mobility who often cannot explore or use certain environments without proper accommodations. According to the CDC, 12.2\% American adults have a mobility disability \cite{CDCDisabilityMobility2019}, with a majority of people expected to face mobility challenges as they age. In response, significant efforts have been made to identify accessibility concerns to renovate or build more accessible physical environments and inform people with limited mobility about potential challenges in unknown places \cite{costanza2020design, switzer2003disabled, zallioInclusionDiversityEquity2021, preiserDesignInterventionRoutledge2015, lindenAccessibilityExperienceOpportunities2016, fletcherChallengeInclusiveDesign2015}. 

To scalably identify accessibility concerns, accessibility practitioners have codified common accessibility problems into standardized accessibility checklists like the Home Safety Self-Assessment Tool \cite{horowitzUseHomeSafety2016} and ADA building codes \cite{ADAChecklist2024}. These checklists allow non-experts to evaluate and enforce the accessibility of environments more easily. Researchers have expanded upon this approach through automated accessibility auditing tools like RASSAR \cite{su2024rassar}, which uses mobile AR and computer vision to identify pre-defined accessibility features and crowdsourcing platforms like Project Sidewalk  \cite{saha2019project}, which defines a set of key sidewalk accessibility features for crowdworkers to annotate. While these systems are a cost effective and easily scalable way to collect large amounts of data, their checklist-based approach to identifying accesibility accessibility concerns fail to consider an individual's unique abilities, needs, preferences and how they change over time. Thus, the accessibility information generated from these approaches fail to capture how people personally experience accessibility in their physical environment \cite{curlSameQuestionDifferent2015, lattmanNewApproachAccessibility2018, mccormackObjectivePerceivedWalking2008, vandervlugtWhatPeopleDeveloping2019, potPerceivedAccessibilityWhat2021, kasemsuppakornUnderstandingRouteChoices2015, kirkProblemsGeography1963, morrisAccessibilityIndicatorsTransport1979}. This mismatch can create a misleading perception of accessibility, potentially leading people with disabilities into frustrating and dangerous situations or imposing needless restrictions that further limit spatial opportunities.


In response, we propose a novel accessibility auditing approach leveraging recent advancements in large language models (LLMs), enabling personalized accessibility scans of built environments at scale. We present \systemname, an LLM-based personalized accessibility assessment system to semi-automatically identify accessibility concerns from images. \systemname accepts images of built environments to identify and visualize personalized accessibility concerns, enabling users to analyze thousands of environments at scale using data readily available from sites like Yelp or Google Maps. Through collaborative annotations, our system allows people with disabilities that affect their mobility to validate outputs from LLMs while incrementally learning their motor capabilities, preferences, and environmental interests to continuously updating its user model and improve its assessments over time.

We first ground our work by conducting a formative study to identify the current difficulties, needs, and process of finding accessible spaces. 
These insights informed the design of \systemname, aligning its prediction pipeline with how users naturally evaluate accessibility concerns and enabling users to guide the system through collaborative human-AI annotations to build a dynamic model of the user's needs and preferences.
We then recruited 10 participants with varying levels of self-described mobility to build personalized user models through hour-long interactions with \systemname and conducted a technical evaluation using generated personalized accessibility annotations across 500 images of built environments. 
Finally, we conducted a user study where participants evaluated the usefulness of personalized versus non-personalized system outputs and shared their thoughts through interviews. These evaluations yielded both quantitative and qualitative insights about our system. Our findings indicate that \systemname effectively generates useful data and engages users, while raising important considerations for future AI systems for accessibility scans.

Our contributions are threefold: First, we introduce the first LLM-based approach to accessibility auditing, enabling semi-automatic, personalized assessments. We show that scalable personalization is possible using low-cost inference methods and readily available online data. Third, we provide novel insights from our user-centered design and evaluation, highlighting effective LLM user modeling strategies, new accessibility scanning methods, and new interaction techniques for human-AI collaborative annotations in accessibility. We believe \systemname paves a new path toward personalized accessibility assessment, with the potential to transform current accessibility practices. This potential is amplified by the vast number of online images (\eg, from Airbnb to Yelp), which \systemname could use to support people with limited mobility in making travel decisions, room reservations, and event planning.

\section{Related Work}
We situate our work in digital accessibility assessment, visual affordance predictions, LLM-based personalization, and mobility modeling for accessibility.

\subsection{Digital Accessibility Assessment}
Recent innovations in sensing and computing have advanced environmental accessibility assessments. For example, Bring Environments to People \cite{chi2023bring} lets people with limited mobility remotely assess spaces through browser-based virtual tours. Other research \cite{peiEmbodied2023, harrisonDevelopmentWheelchairVirtual2000, kaklanisVirtualUserModels2013, perezWUADWheelchairUser2022, moussaouiVirtualRealityAccessibility2012} builds upon this idea by utilizing embodiment techniques in virtual reality (VR), allowing users to explore and assess digital twins of physical environments. Crowdsourced approaches like Project Sidewalk \cite{saha2019project, sharifExperimentalCrowd+AIApproaches2021, weldDeepLearningAutomatically2019} use online crowdworkers to remotely label accessibility issues. Researchers have also attempted to automate accessibility auditing using computer vision to capture precise measurements and detect key issues in built environments using smartphones \cite{su2024rassar}, robotic mechanisms \cite{su2024rais, su2024demo}, existing imagery from Google Earth and Google Street View \cite {itoAssessingBikeabilityStreet2021, seekinsExploringEnvironmentalMeasures2022}, 3D scenes \cite{fuHumancentricMetricsIndoor2020}, scene graphs \cite{daoThreedimensionalIndoorNetwork2018, fuHumancentricMetricsIndoor2020}, and point clouds \cite{sernaUrbanAccessibilityDiagnosis2013, baladoAutomaticBuildingAccessibility2017, anjanappaDeepLearning3D2022}.
However, current digital accessibility assessment solutions generally lack sufficient customizability and personalization features to support the diverse needs of the disability community. \systemname addresses this issue by using LLMs and human-AI collaborations to continuously model a user's needs and generate more personalized accessibility scans.


\revised{
\subsection{Visual Affordance Prediction}
Evaluating accessibility through images can be viewed as an application of visual affordance prediction, the process of using visual cues to identify how an object should be used. Recent research have shown how we can infer affordance from images through the innate properties of objects \cite{yaoDiscoveringObjectFunctionality2013, chaoMiningSemanticAffordances2015}, physical and social boundaries in a scene, \cite{chuangLearningActProperly2018}, and specific interactions like grabbing \cite{songLearningDetectVisual2016}. Affordance prediction has become especially important with the rise of semi-autonomous robots, where robots must first identify whether a certain action is feasible before attempting it \cite{besslerdanielFormalModelAffordances2020, minAffordanceResearchDevelopmental2016, chengEmpoweringLargeLanguage2024}. This process is similar to accessibility evaluations, where users must identify the feasibility of completing specific actions before traveling to the location. Closer to our work integrating recent developments in LLMs, recent developments in computer vision demonstrate how LLMs can be grounded to produce better affordance predictions \cite{ qianAffordanceLLMGroundingAffordance2024, chenWorldAffordAffordanceGrounding2024} and how LLMs can leverage visual affordance to improve outputs \cite {liImprovingVisionLanguageActionModels2024, chengEmpoweringLargeLanguage2024}. We view \systemname as a human-centered approach to LLM-based visual affordance approaches where users leverage chain-of-thought prompting techniques to guide LLMs to consider specific tasks and that task's affordance in relation to the user's capabilities and environmental features. 
}

\subsection{LLM-based Personalization}
Interest in LLM-based personalization is growing across various domains \cite{wozniak2024personalized,zhang2024personalization, mamykina2022grand, zhangPersonalizationLargeLanguage2024}. Recent research has used LLMs to simulate human test subject responses in Turing tests \cite{aher2023using} and replicate individual attitudes and behavior in interview responses \cite{park2024generative, hamalainen2023evaluating}. LLM agents have also been used to simulate user feedback on user interface usability \cite{xiang2024simuser, duan2024generating}. Harrak \textit{et al.} \cite{benharrak2024writer} used LLM-based personalization to synthesize on-demand feedback from a target audience with LLM-based personas.

Closer to our work using LLMs to generate accessibility insights that influence how users choose environments are LLM-based personalized recommendation systems. Researchers have used LLMs to capture preferences of blind and low-vision individuals for navigational aid \cite{an2025can}. Joko \textit{et al.} \cite{jokoDoingPersonalLAPS2024} demonstrated the use of LLMs to guide the creation of more aligned user preferences that better match users' actual preferences. Other research explored how LLMs can democratize personal health insights \cite{merrillTransformingWearableData2024, cosentinoPersonalHealthLarge2024} and create personal medical assistants \cite{zhangLLMbasedMedicalAssistant2024}. LLMs have also been used to enhance traditional recommendation algorithms. Zhang \textit{et al.} demonstrates the addition of smartphone sensory data to improve the emotional response of recommendations through LLMs. Other works demonstrate that the integration of LLMs into existing systems can directly improve recommendation quality \cite{baoTALLRecEffectiveEfficient2023, lyuLLMRecPersonalizedRecommendation2024, liuONCEBoostingContentbased2023}.

\subsection{Mobility Modeling for Accessibility}

HCI reserachers have modeled motor capabilities to evaluate ergonomics \cite{lamkullInfluenceVirtualHuman2007,kurschlUserModelingPeople2014, schalljrDigitalHumanModeling2018} and assess the usability of different technologies and systems for a variety of audiences. While these user modeling frameworks were designed for general populations, other works focus primarily on people with limited mobility. Huang \textit{et al.} \cite{huang2024wheelpose} improves pose estimation for wheelchair users using synthetic data from photorealistic avatars driven by user-centered motion generation techniques. More closely related is literature on accessibility simulations using virtual human agents. Kaklanis \textit{et al.} \cite{kaklanisVirtualUserModels2013} models older adults and people with disabilities by breaking down tasks into hierachical motions and interactions for ergonomics testing of product prototypes in different scenarios and tasks. \textit{Embodied Exploration} \cite{peiEmbodied2023} models wheelchair users with avatars in VR using three key dimensional parameters: wheelchair maximum width, wheelchair armrest height, and seated eye height. These parameters are used to enhance the embodied experience, enabling more accurate accessibility assessments in VR environments. More automated utilization of user models include recent works in transportation network accessibility evaluations, using GIS-based networks \cite{metcalfModelingSocialDimensions2013, zhangModelingAccessibilityScreening2018} to model how user preferences and capabilities might affect route finding.

\revised{
\systemname introduces a dynamic new approach to digital accessibility assessment by harnessing recent advances in LLM-based collaborative annotation, visual affordance prediction, and personalization. Unlike traditional methods, \systemname empowers users to construct individualized user models that adapt to their specific needs and preferences, enabling image-based accessibility evaluations that more accurately reflect how they would experience and navigate a given environment before travel.
}

\label{sec:needfinding}
\section{Formative Study}


To inform the design of an LLM-based accessibility assessment system, we performed an initial formative study with the following goals: (1) to advance understanding of the current practices and challenges of accessibility assessment for people with limited mobility; (2) to investigate the feasibility of using LLMs for accessibility assessments; (3) to elicit user feedback on the idea of personalized accessibility scans. We recruited six participants (U1-U6), all self-identified as daily wheelchair users and were compensated with $\$20$ for their time. Refer to Appendix \Cref{tab:needfinding-demographics} for demographics. 

\begin{figure}
    \centering
    \includegraphics[width=1\linewidth]{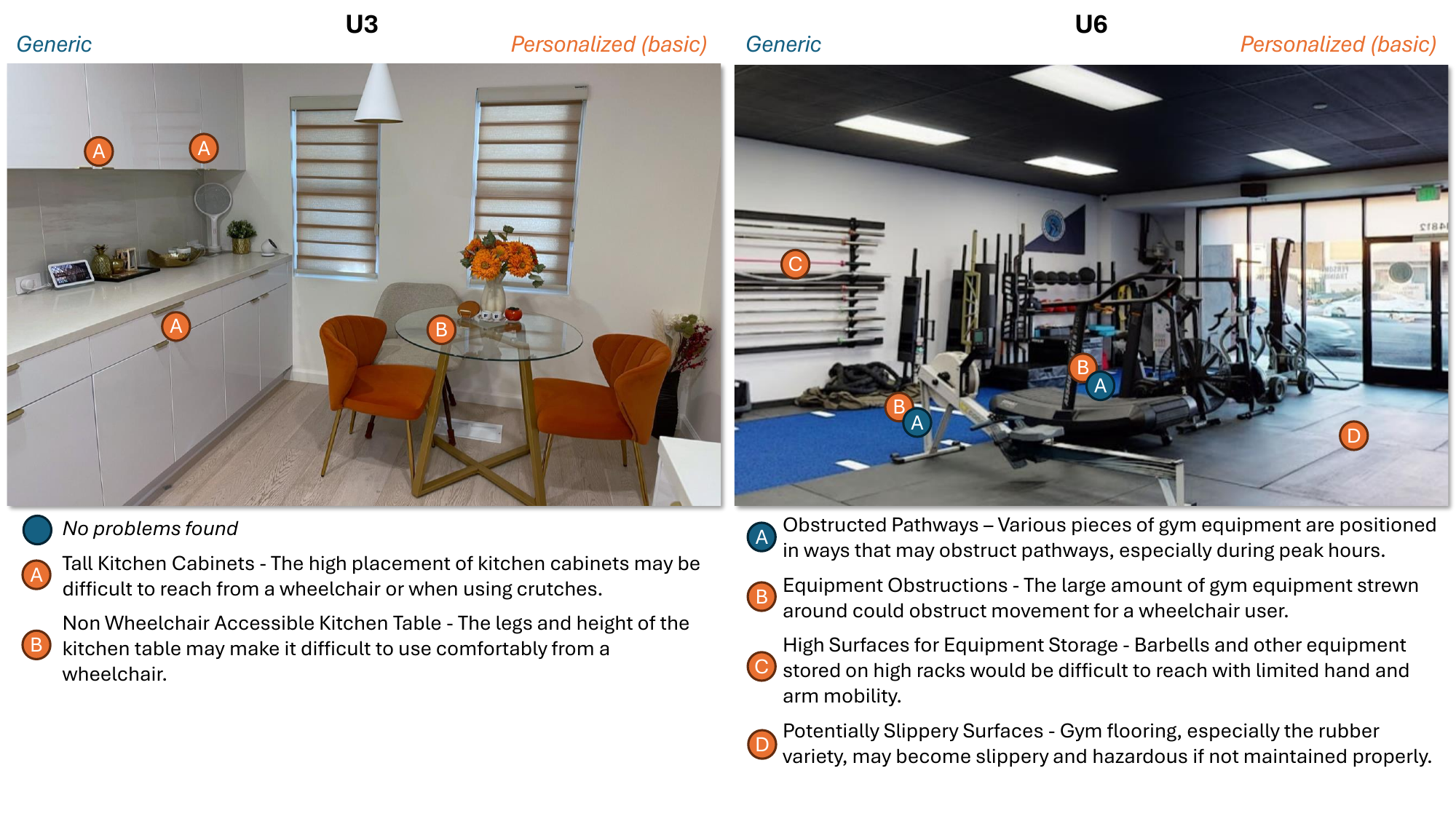}
    \vspace{-0.8cm}
    \caption{Two examples that show the comparisons between generic LLM-generated annotations and those with basic personalization for two participants in our formative study. }
    \Description{
    Comparison of generic and personalized LLM-generated accessibility annotations for two users' environments in formative study. Left: U3's kitchen with two personalized annotations highlighting cabinet height and table inaccessibility for wheelchair use. Right: U6's gym with four personalized annotations pointing out obstructed pathways, scattered equipment, high storage racks, and slippery flooring.
    }
    \label{fig:needfinding-example}
    \vspace{-0.6cm}
\end{figure}

\subsection{Procedure and Analysis}
Before the study, participants were sent an initial survey about demographics and self-described motor capabilities. Two researchers then conducted a \char`\~30-min virtual research study with each participant individually through online meetings. \revised{Our study was approved by our institution's IRB.}

Each study began with a semi-structured interview where participants were asked about their current experience virtually analyzing environments to further understand the needs and challenges of existing solutions. Participants were then shown images of different places and asked to think-aloud on how they would assess the accessibility of the environment. Researchers then generated two sets of accessibility concerns using OpenAI GPT-4o-2024-08-06 \cite{HelloGPT4o} by prompting \textit{with} and \textit{without} the user's self-described motor capabilities listed in \Cref{tab:needfinding-demographics}. Researchers then visually annotated each concern on the image and asked participants which set of annotations they preferred.

We collected audio recordings and notes, which were analyzed via thematic analysis \cite{braunUsingThematicAnalysis2006} by two researchers to identify key themes and concerns. The first researcher, who also conducted the interviews, reviewed all transcripts and notes to develop an initial codebook. The two researchers then discussed the codebook, iteratively resolved disagreements, and developed a final version of the codebook. The first researcher then converted this codebook into identified themes. Participant quotes have been lightly edited for concision, grammar, and anonymity.

\subsection{Findings}
We highlight four key findings below in regards to existing accessibility assessment practices, personalization in accessibility, and reactions to LLMs for accessibility assessment:

\textbf{Existing accessibility assessments deter exploring new environments.}
All participants agreed that current methods for environmental evaluation were difficult or insufficient. Many participants cited this as a reason why they do not explore new environments with comments like \textit{"even though I've done my due diligence to ensure accessibility to my needs, there's a reasonable chance that I'll get there, and it's still going to **** up."} (U6). The tedious process of finding data can be a deterrent in and of itself, with U5 stating that \textit{"I know we might want to go try a new place and with having to Google [accessibility]. I might just say, no. Let's just go somewhere where we know it's going to be accessible. So that is definitely a deterrent"} (U5).

\textbf{Merits of LLMs for automatic accessibility assessments.}
All participants stated that the LLM annotations with and without their self-provided information were useful, especially when compared to existing accessibility data available online. Researchers also note that all participants found that the LLM's generated accessibility scans were equal or improved when prompted with the participant's self-described mobility. Example feedback regarding the usage of the LLM included \textit{"It's amazing! It's not only wonderful information, it makes you feel more included too"} (U3); \textit{"This is, gonna be so useful. It's going to be helpful even."} (U5); \textit{"Because [the LLM] will identify restaurants or hotels that will address my needs"} (U4).

\textbf{Need for personalized accessibility assessments.}
Throughout the study, participants appreciated the personalized accessibility assessments as seen in \Cref{fig:needfinding-example}. Researchers also noted that all participants evaluated environments differently during the think-aloud exercise, focusing on different aspects depending on their unique needs. During the annotations evaluations, U3 appreciated the personalized annotations mentioning their specific needs, stating \textit{"It mentions that the navigating with wheelchair or crutches. That's beautiful. You don't see that a lot. Usually it just concentrates just on wheelchairs. So that is so awesome that that's mentioned."} (U3). U6 pointed out that his assessment of accessibility depends on the task at hand where \textit{"The intended function. Is it going to be my permanent home? Is it going to be a temporary residence. Yeah, it absolutely changes my requirements of the space"} (U6).

\textbf{Supportive features to improve usability.}
Researchers found that participants varied in the amount of detail on accessibility concerns they preferred. U1 points out that when they are conducting research on environments, \textit{"I'm looking for a particular thing. If I wanted more maybe I could click to get more, but the main points are just simple and quick."} (U1). U5 requested more detail stating, \textit{"There's different levels of wheelchair accessibility. So that extra detail is super helpful."} (U5). Participants also valued the use of visual markers to indicate accessibility concerns with comments like \textit{"I could see how the [visual annotations] would be very useful, because [other people] are not in a wheelchair. They don't see those things that I do."} (U5). U6 also suggested that the system could be extended further to automatically handle environmental inquiries, \textit{"I think you could do the calling in, if not with a you know actual voice, certainly by automated emails."} (U6).

\begin{figure*}
    \centering
    \includegraphics[width=1\linewidth]{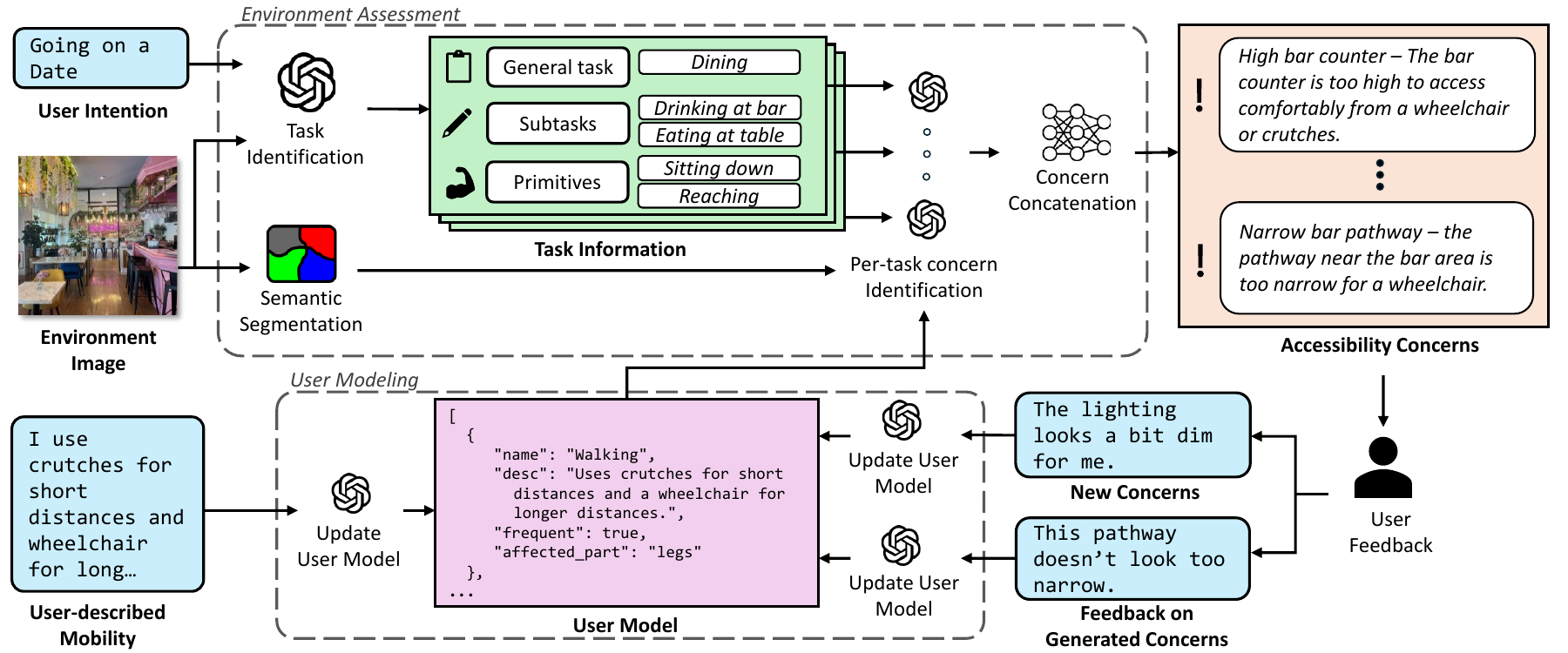}
    \caption{ \systemname system structure. Users first submit an image of an environment and their intent. \systemname identifies potential tasks and their actions to predict environment concerns. The user can then provide feedback on these concerns to update their user model, which leads to improved accessibility assessment in future scans.}
    \label{fig:system-diagram}
    \Description{
    Two main components: Environment assessment and User modeling. Environment assessment takes in an image of an environment and user intention. Environment assessment consists of task identification and semantic segmentation of images which generates task hierarchies (general tasks, subtasks, and action primitives). Task hierarchies are then used for per-task concern identification and concern concatenation to produce accessibility concerns. Outputted accessibility concerns can be used with human feedback to feed into User modeling which stores and updates user model profiles. Feedback loops from user concerns refine model and concern predictions to improve future assessments.
    }
\end{figure*}

\section{System Design and Implementation}

Following findings from our formative study showing how simple LLM prompts can produce useful accessibility scans, we developed \systemname, which combines LLMs and computer vision to model a user's accessibility preferences and enable semi-automatic, personalized, accessibility scans. All system components were developed using OpenAI ChatGPT-4o-2024-08-06 \cite{HelloGPT4o}. See \Cref{fig:system-diagram} for a system diagram.

\subsection{Design Considerations}
We developed \systemname with the primary objective of enabling users to run accessibility scans personalized to their individual needs at scale. We report on the following design considerations derived from our formative study:

\textbf{D1: Support context-aware and adaptive personalization.}
Participants in our formative study often qualified their assessment of accessibility by the specific times and tasks they engage in. This goal aligns with findings from Lättman \etal's \cite{lattmanNewApproachAccessibility2018}, which detailed how perceived accessibility is highly context-dependent. 
Participants also emphasized that their accessibility needs change over time, whether due to physical decline or new techniques to navigate previous inaccessibilities. Therefore, a robust accessibility scanning system should support adaptive personalization by allowing users to provide feedback to address evolving needs and situational contexts. 

\textbf{D2: Enable human oversight over AI.} 
While initial findings from our formative study indicate LLMs are capable of generating useful accessibility scans, we must acknowledge their susceptibility to inaccuracies and hallucinations. Participants echoed these sentiments, emphasizing the need for a tool that works \textit{with} them and not \textit{for} them. To support this, users should be able to review and revise LLM-generated annotations, facilitating both personalization and error correction. Our system should also be easily interpretable, helping users to easily understand and guide the scanning process. 

\textbf{D3: Generate detailed accessibility concern information.} 
Participants preferred varying levels of detail in accessibility information depending on how important it was to access a given space. They also highlighted the value of being able to explore additional details about a concern when needed. \systemname should support these needs by providing a variety of accessibility information, allowing users to quickly glance at key insights or conduct a more in-depth analysis of the built environment as needed. 

\textbf{D4: Enable the system to run at scale.} 
Participants in our formative study were especially excited about using a potential LLM-based system to more efficiently discover accessible environments. For example, participants envisioned using such a system to evaluate multiple Airbnbs more efficiently than would be possible through manual inspection, simplifying the search for accessible vacation options. To support such use cases, \systemname must be able to scale to handle large volumes of built environments.

\subsection{User Modeling}
\systemname's user modeling component handles the maintenance of a structured user model which can be iteratively updated through user feedback and is easily interpreted by LLMs.

\textbf{User model structure. }
To facilitate both human interpretability and AI performance, we represent the user model in JavaScript Object Notation (JSON) format. Each user model consists of a set of attributes which describes a specific movement (\eg, reaching above with my right arm), how the movement might be affected (\eg, I can not reach above shoulder level), whether the movement is frequently performed, and the affected body part or preference (arms, legs, feet back, chest, hands, eyes, ears, brain, user preference). An example of a user model attribute is shown in \Cref{fig:system-diagram}. This user model structure can capture physical attributes of a user (\eg, footprint of a wheelchair), sensory and cognitive attributes (\eg, sensitivity to sound), and the user's value system (\eg, prefer quieter places) which has been shown to better represent a user's decision making \cite{potPerceivedAccessibilityWhat2021}, and provides additional context to the importance of specific subtasks through the frequency boolean. Furthermore, JSON attributes allow various details, including context and specific scenarios, which can grow and shrink over time.


\textbf{Elicitation methods. }
To generate the user model, we enable three different elicitation methods: (1) \textit{Self-Description.} Users are able to enter an unstructured textual description of their capabilities and preferences which can include recounts of prior experiences. An LLM then decomposes this self description into a series of affected motions which can be input as the user model. (2) \textit{Environmental Annotations.} Users can also choose to annotate concerns in images of different environments. The concerns, their reasoning, and the image are entered into an LLM to generate a user model. (3) \textit{Feedback on Environmental Annotations.} Users can also update their user model by providing feedback to AI-generated environmental annotations. User feedback, the original annotation, and the image are then inputted into an LLM and used to update the user model. All elicitation methods are implemented through textual input. User interfaces to utilize elicitation methods are detailed in \Cref{sec:user-interface}.

\begin{figure*}
    \centering
    \includegraphics[width=1\linewidth]{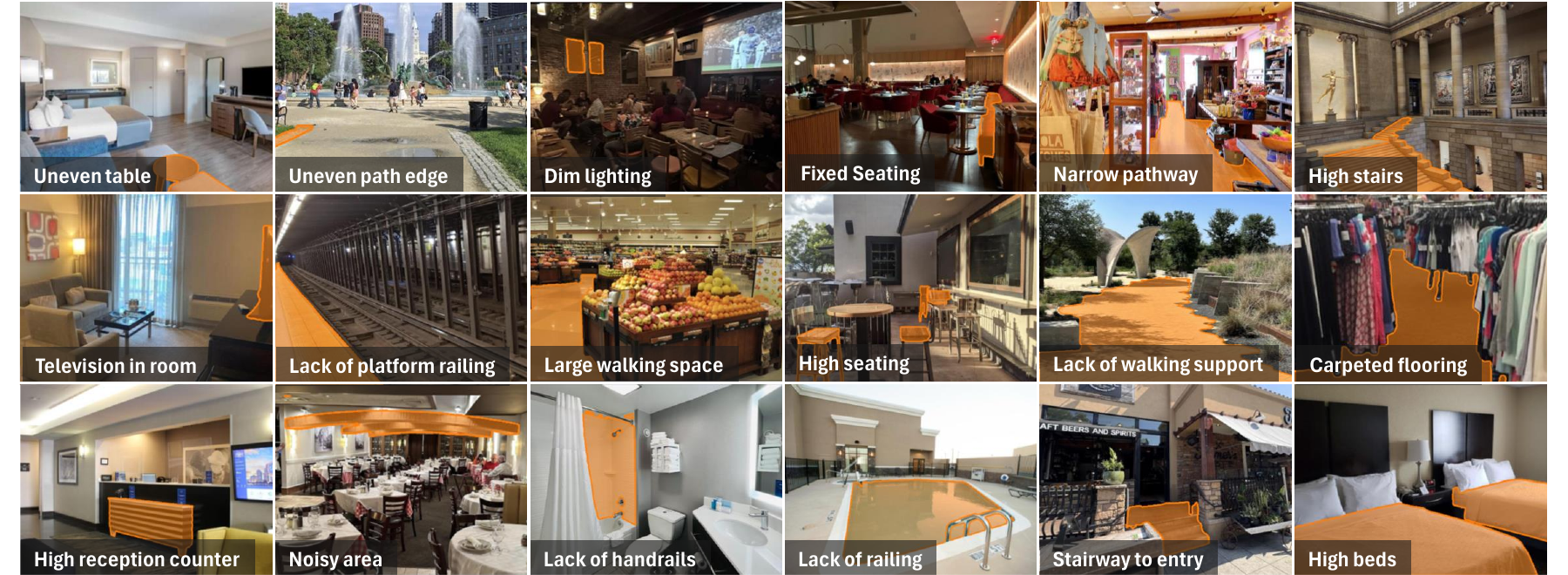}
    \caption{Example accessibility concerns identified by \systemname. These concerns range from narrow floorplans and furniture height to general facets of environment like the presence of specific objects or noise. Note: texts shown are summarizations from detailed concerns. Full unabbreviated examples are in \Cref{fig:teaser}.}
    \Description{Grid of images showing diverse accessibility concerns detected by Accessibility Scout. Detected concerns includr uneven tables, high seating, narrow pathways, dim lighting, noisy areas, high stairs, lack of railings or handrails, high beds, and other environmental or layout-related barriers.}
    \label{fig:example-outputs}
\end{figure*}

\begin{figure}[t]
    \centering
    \includegraphics[width=1\linewidth]{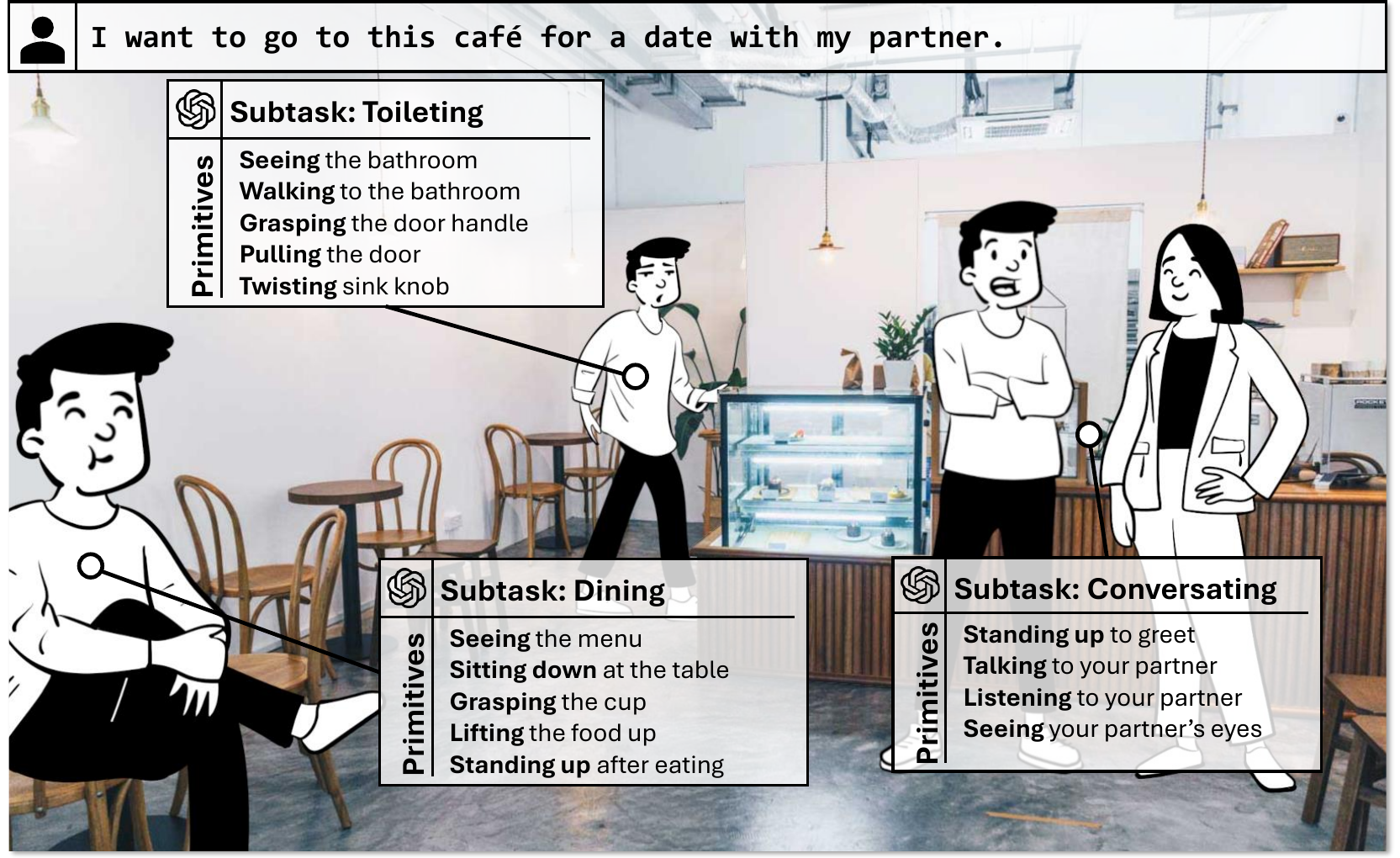}
    \caption{Visualization of how tasks are broken down by \systemname. Given a desired user intent like "Going on a date", the system first breaks down the task into subtasks the user might perform during the task like "toileting", "conversating", or "dining" and potential locations for these subtasks. Upon doing so, the system then further breaks down these subtasks into primitive motions.}
    \Description{Visualization of task breakdown featuring three subtasks: toileting, dining, and conversating. Tasks are derived from the user intent "Going to this café for a date." Each subtask is linked to a character in the café image and broken into primitive motions. Toileting includes actions like walking to the bathroom and twisting a sink knob; Dining includes sitting, eating, and lifting food; Conversating includes standing up to greet and making eye contact.}
    \label{fig:task-breakdown}
\end{figure}

\subsection{Accessibility Assessment}
\systemname uses the user model to generate accessibility concerns on images of different environments (\Cref{fig:system-diagram}). In order to build a more human-interpretable and controllable system, we designed the assessment process to mimic how participants from our formative study evaluated the accessibility of places by the tasks they may prohibit. This serves three primary purposes: (1) To address the need for task-specific accessibility assessments identified through our formative study. (2) To generate more relevant and interesting concerns. (3) To enable parallelized LLM requests for scalability and reliability. 

\textbf{Task identification. }
To generate comprehensive and detailed accessibility scans, we first identify the spatial tasks a user could engage in based on their intended use of the environment. Users first input an image of an environment and a short description of the environment and their intended usage into an LLM which is prompted to predict a set of common tasks that a user might perform in the environment (\eg, study at a cafe). \revised{Within the same context window, the LLM is then prompted to decompose each one of these tasks into subtasks  (\eg reading your textbook is a subtask of studying at a cafe). Each subtask consists of a short description, potential locations these subtasks might be performed in an environment, and primitive motions necessary, a concept derived from Kaklanis et al.\cite{kaklanisVirtualUserModels2013} which models any task as a series of fundamental movements like grabbing, reaching, and pulling. \Cref{fig:task-breakdown} illustrates an example of the identified tasks and fundamental motions required when a user specifies their intended use of the environment for a date. By first breaking down the environment into potential tasks and subtasks, we treat accessibility assessments as a task affordance problem which is both more reflective to what a user might actually need and enables more focused LLM contexts to generate better predictions later on.} Prompts for are shown in Appendix \Cref{fig:appendix-prompts-identify-tasks} and \Cref{fig:appendix-prompts-identify-primitives}.

\begin{figure*}[h]
    \centering
    \includegraphics[width=1\linewidth]{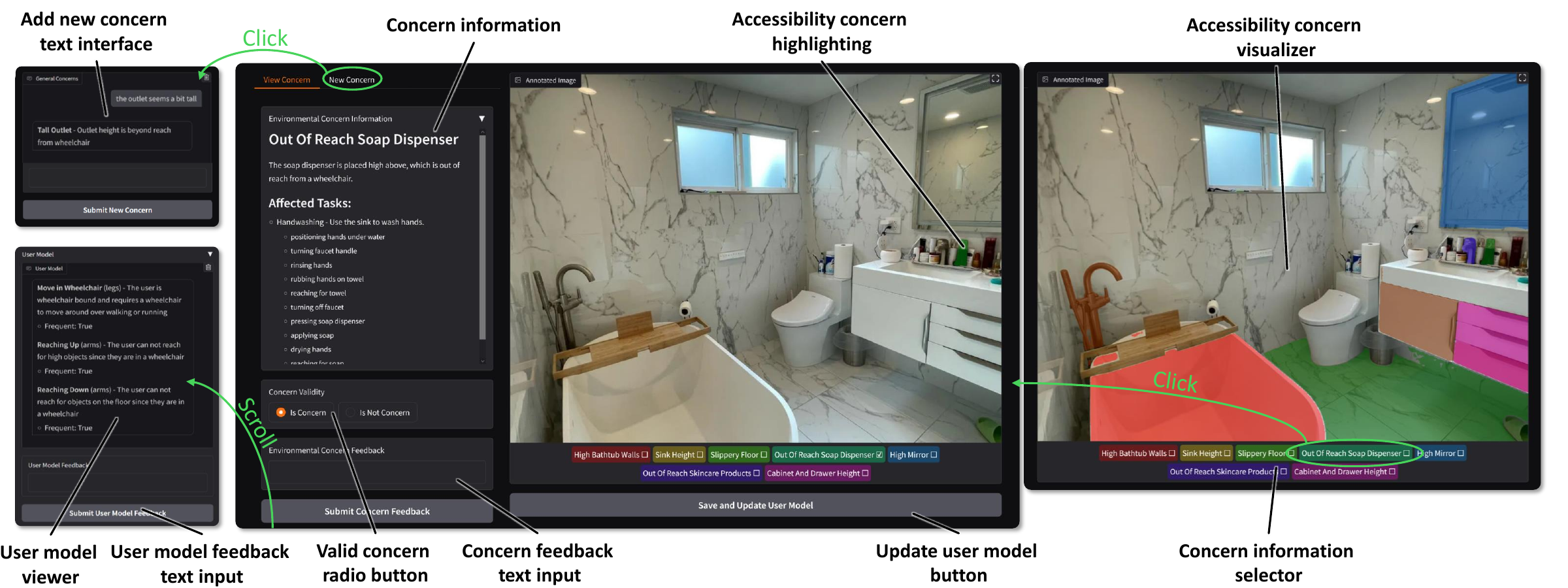}
    \caption{The \systemname UI. Right: The user can view all detected concerns in the visualizer. Middle: Upon selecting a specific concern, information on the concern and highlighting are then shown. The user is then able to provide feedback on the concern to improve the user model. Top left: The text interface for users to add more concerns not identified already. Bottom left: The text interface for users to view their user model and provide unstructured text feedback.}
    \Description{Four interface components of Accessibility Scout. Right: Visualizer shows annotated bathroom with color-coded accessibility concerns. Clicking on a concern opens the concern details in the middle. Middle: Concern selection displays related text details and tasks affected, with options for marking concern validity and submitting feedback. Top left: Text box for users to add new concerns not automatically detected. Accessed from the "New Concern" button. Bottom left: Viewer shows user model and a feedback text area for refining it. }
    \label{fig:ui-flow}
\end{figure*}

\textbf{Concern identification. }
To identify key environmental concerns, we use Set-of-Mark Prompting \cite{yangSetMarkPromptingUnleashes2023}. Images are overlaid with semantic segmentation masks generated from Semantic-SAM \cite{liSemanticSAMSegmentRecognize2023}, providing textual labels for different semantic segmentation masks of the image to enhance the LLM's spatial understanding. Masks are later used for visualization purposes in the user interface. \revised{For each task generated from the task identification process, its task description, list of subtasks, the user model, and Set-of-Mark prompts are fed into an LLM prompted to identify key parts of the environment that would prohibit the task.} For each task, the LLM outputs a set of environmental concerns. Each concern consists of a short name, reason for why the concern was identified for the user, and the location by the label generated from Semantic-SAM segmentation. We note these concerns are qualitative heuristics like \textit{"low"} or \textit{"soft"} and not precise measurements. Each task is processed in a parallel LLM request for speed, smaller context and more focused context windows, and partial outputs in cases where the API endpoint is unstable. Prompts for this process are shown in Appendix \Cref{fig:appendix-prompts-identify-concerns}.

\textbf{Concern concatenation.}
Since concerns are generated in parallel, requests can generate redundant accessibility concerns.  Thus, all concerns are grouped through semantic text similarity analysis using Sentence-BERT \texttt{all-MiniLM-L6-v2} \cite{reimers-2019-sentence-bert} with the name and reason of each concern. Concerns with a cosine similarity over a threshold of $0.7$, selected through experimental analysis, are combined by selecting the name and reasoning with the highest average cosine similarity in the group (\eg, \textit{High Bar Counter - The bar counter is too high for the user to access comfortably from a wheelchair} and \textit{High Bar Counter - The height of the bar counter makes it difficult for the user to reach drinks or interact comfortably}). Examples of final accessibility concerns are illustrated in \Cref{fig:teaser}, highlighting various issues identified for users with different mobility levels and preferences. \Cref{fig:example-outputs} presents a broader range of compatible photos of built environments.

\label{sec:user-interface}
\subsection{User Interfaces}
\systemname has a web user interface implemented in Gradio \cite{abidGradioHassleFreeSharing2019} that takes images of environments and generates indicators on potential accessibility concerns which users can provide feedback on (\Cref{fig:ui-flow}). The following describes an example user scenario: 

Alexandria is a wheelchair user using \systemname to evaluate the accessibility of an Airbnb. Upon starting \systemname with a picture of the Airbnb's bathroom and her user model, Alexandria is greeted with a visualization of all detected concerns (\Cref{fig:ui-flow} right). She hovers her cursor over the selector labeled \textit{Out of Reach Soap Dispenser} which highlights the concern in the visualizer. She then clicks on the selector which changes the accessibility information textbox to show that the soap dispenser is inaccessible because it is too high on the counter (\Cref{fig:ui-flow} middle). Alexandria agrees with this generated concern, selecting the \textit{Is Concern} radio button, and provides further feedback in the \textit{Envrionmental Concern Feedback} textbox that \textit{"The soap dispenser is just too far back on the counter to reach comfortably"}. Alexandria repeats this process for all other identified concerns before noticing that she might not be able to plug in a hairdryer since the outlet is high up. She clicks the \textit{New Concern} button and types in the text interface: \textit{"The outlet seems a bit tall"} (\Cref{fig:ui-flow} top left). She presses enter (same as clicking on the \textit{Submit New Concern} button) and the text interface and visualizer show a new concern: \textit{Tall Outlet}. She scrolls down to the user model viewer and sees a new attribute: \textit{Outlet height is beyond reach from wheelchair} (\Cref{fig:ui-flow} bottom left). She finally finishes evaluating all generated concerns and clicks \textit{Save and Update User Model} which prompts \systemname to update her user model with her feedback on this image.

\label{sec:tech-eval}
\section{Technical Validation}
We investigate the feasibility of using LLMs to assess the personalized accessibility of environments and ground our findings from the subsequent user study through a set of technical evaluations of \systemname in 500 images of different environments using varying user models. Through our technical evaluations, we note that it is unfeasible to directly evaluate the quality of identified concern heuristics given their accuracy is subjective (\eg, a table of 34 inches might be okay for someone in a manual wheelchair to sit at but too low for a power wheelchair). Instead, we defer evaluations on usefulness to our subsequent user study. In this section, we focus on evaluating the following properties of \systemname: (1) Accurate detection of environmental features as a measure of hallucinations. (2) Distribution of identified concerns as a measure of \systemname's ability to capture accessibility needs in different environments. (3) Differences in detected concerns between user models as a measure of degree of personalization. (4) Cost of evaluation as a measure of scalability.

\subsection{Data Collection}
\label{subsec:tech-data-setup}

Researchers compiled 500 images of different environments across 9 of the most populous cities in each region of the United States (Los Angeles, San Diego New York City, Philadelphia, Chicago, Columbus, Phoenix, San Antonio, Houston) sourced from searches through Google Maps and Yelp. Eight crowdworkers first used keywords of commonly accessed environments (community centers, grocery stores, lodging, restaurants, retail stores, transportation hubs, and public venues) across North America to assemble an initial dataset of images of environments. Researchers then manually evaluated and selected 500 images according to the following criteria: Images showed key parts of the environment including pathways, utilities, restrooms, and functional areas necessary to complete the purpose of the area. Images had a wide enough field of view to capture the entire environment (\eg full view of the floor up). All images were then briefly labeled with a general description of what someone might be doing in that environment. Examples from the dataset are shown in \Cref{fig:example-outputs}.
Using \systemname, we run an accessibility scan of each image using each of the user models created by participants in the first stage of a later user study, which will be detailed in \Cref{sec:user-study}, as well as an empty \textit{"generic"} user model for a total of 11 user models. Demographic information used to initialize \systemname are shown in \Cref{tab:user-study-demographics}. As a result, we generate 5,500 accessibility scans (39,394 concerns).

\begin{table}[t]
\caption{Demographic and self-described mobility  of 10 participants (P1-P10) in the user study.}
\label{tab:user-study-demographics}
\scriptsize
\begin{tabular}{@{}p{.06cm}p{.06cm}p{.35cm}p{1cm}p{5.5cm}@{}}
\toprule
ID & Age & Gender & Diagnosed disability & Self-described mobility and preferences \\ \midrule
P1 & 58 & F & Spinal cord injury, Fibromyalgia, Stenosis, Arthritis & Use a manual wheelchair with attached motor. Able to stand and pivot for transfers. Upper extremities are weak with limited strength and dexterity. Visual and hearing difficulties. \\
P2 & 54 & M & Spinal cord injury (T11/T12, L4/L5) & Crutches for short distance and wheelchair for long distances. Not injured from waist up and have full function in chest, arms, hands, shoulders, and neck. Nerve damage in hips, hamstring, and no function from knee down. \\
P3 & 61 & F & Polio, Post-polio syndrome & Ambulation causes physical and mental strain due to tripping risk. Stairs, inclines, uneven surfaces, and slick surfaces are difficult. Cannot stoop, rise off floor, or rise from low toilet without support. Stepping in/out of tubs is dangerous but possible with grab bars. High bar stools are difficult. Holding anything in one hand can be difficult as I walk with a forearm crutch. Long distances require a rollator. Muscle fatigue requires frequent breaks. \\
P4 & 72 & F & No diagnosed disability & Cannot walk extended distances due to fatigue. Long distances, staircases are problematic. Handrails are important. \\
P5 & 86 & F & Post-polio syndrome & Need to use a walker. Right leg is super weak, especially knee. Have a hyper extended knee and drop foot. Use a knee brace and ankle foot orthosis. Cannot walk long distances. \\
P6 & 35 & M & Spinal cord injury (T10) & Use a manual wheelchair. Prefer rolling on hard floors. Carpet is difficult to push. Can move pretty comfortably and smoothly in wheelchair and can go anywhere except sandy places. \\
P7 & 58 & M & Quadriplegic (C7) & Use a manual wheelchair and SmartDrive assistive device. Prefer hard floors, wide corridors, low/no thresholds in doorways, automated entry doors. Need restroom facilities that accommodate wheelchairs with designs that follow ADA. Appreciate buildings with elevators that I can operate over special lifts that require assistance. Appreciate using main entrances over special side entrances. \\
P8 & 38 & M & Quadriplegic (C4) & Use an electric wheelchair. Wide hallways and doorways are a must. Prefer hard, smooth floors and no thick carpet. Need elevators with wide automatic doors and enough space inside. Ramps need a gentle slope. Need accessible restrooms with enough space to turn. I do not transfer so toileting and grab bars are not important. Prefer lever-style door handles or automatic door handles since twisting doorknobs is challenging. Paddle switches work best for lighting. Good lighting is a must to see everything. Need to watch for rough or uneven surfaces. Long distances are difficult due to battery constraints. \\
P9 & 66 & M & Paraplegic (T4) & Use a manual wheelchair. Need doors wide enough for wheelchair. Prefer 1 level places and ramps/elevators instead of stairs. Prefer grab bars in rest rooms, hand drying equipment next to sink, hard floors, and doors that swing out. \\
P10 & 50 & M & Quadriplegic & Use a power wheelchair and prefer environments with smooth flat surfaces and open. spaces. \\ \bottomrule
\end{tabular}
\Description{Table listing demographic and mobility information for 10 participants (P1–P10). Columns include participant ID, age, gender, diagnosed disability, and self-described mobility and preferences. Participants report a range of disabilities including spinal cord injury, post-polio syndrome, and various levels of paralysis. Mobility descriptions note use of wheelchairs, crutches, walkers, and specific preferences such as avoiding stairs, needing grab bars, and favoring smooth or hard flooring. Preferences vary based on strength, fatigue, and environment accessibility.}
\end{table}

\begin{figure}[b]
    \centering
    \includegraphics[width=0.9\linewidth]{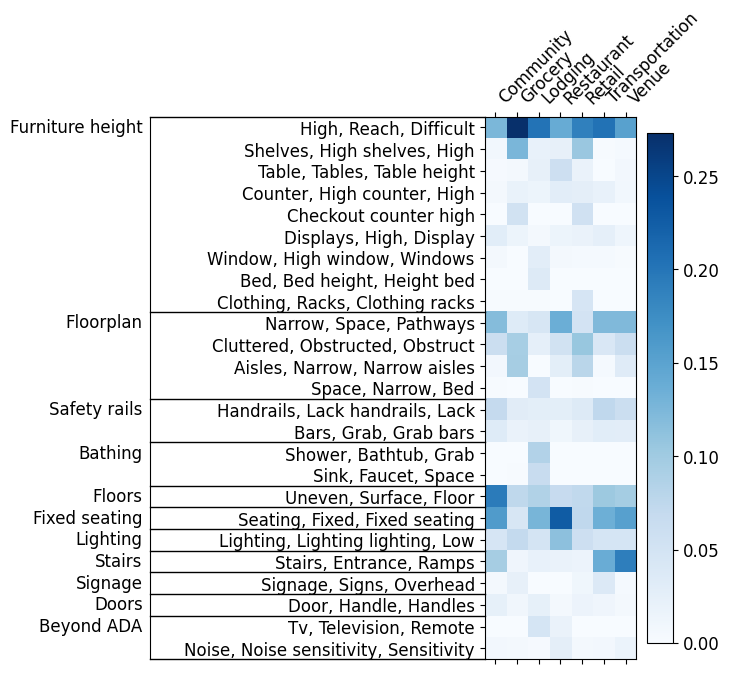}
    \caption{Distribution of twenty-five largest unsupervised clusters of concerns expressed as a percentage of total concerns in each environment type grouped by ADA categories. Labels are three representative keywords for a cluster generated by taking the three terms with the highest \textit{Term Frequency-Inverse Document Frequency}, a measure of a term's importance within a group of text.}
    \label{fig:concern-diversity}
    \Description{Heatmap showing the distribution of 25 largest unsupervised clusters of accessibility concerns across different environment types, grouped by ADA categories. Rows represent concern clusters (e.g., “High, Reach, Difficult” for furniture height), while columns represent environment types (e.g., community, grocery, restaurant). Color intensity reflects the proportion of total concerns in each cluster. Labels use top three TF-IDF terms to summarize each cluster's topic. The darkest cells appear in clusters like "High, Reach, Difficult" for grocery and "Seating, Fixed, Fixed Seating" for restaurants. Less frequent clusters include "Tv, Television, Remote" and "Noise sensitivity." Most clusters show a distribution range of 0–0.15, with few peaking above 0.25.}
\end{figure}

\label{sec:detection-performance}
\subsection{Detection Performance}
Accurate and robust detection forms the foundation of accessibility assessment. \revised{We measure the number of misdectections using a fact-checking approach \cite{huangSurveyHallucinationLarge2025} which only measures object detection accuracy for detected concerns. Since accessibility is highly subjective, we evaluated our system's ability to find useful concerns entirely through our subsequent user study.} Human evaluators manually reviewed all 500 accessibility scans generated by the generic user model, determining whether each identified concern was a hallucination based solely on the following criteria: (1) \textit{Does the related concern exist in the image?} (2) \textit{Does the concern correctly identify the object of concern?} We evaluate purely on this criteria and do not attempt to label object qualities like \textit{"too high"} or \textit{"too soft"} as true or false given users can perceive these qualities differently. Evaluators rated $3590$ concerns and found $237$ ($6.63\%$) hallucinated concerns, which were removed from our later user study. Evaluators noted that hallucinated concerns often centered around specific environmental features not depicted in the image like \textit{checkout counters} or \textit{TV remotes}, potentially indicating that OpenAI-GPT4o has an existing bias towards specific environmental features.

\subsection{System Cost}
We also conducted a basic evaluation on the cost and scanning time of \systemname by measuring the average token usage to evaluate an environment. We compute an average token usage across $500$ images of $8758$ tokens/image ($\text{STD}=1112.175$), $9$ requests/image, and average delay of $10.737$s ($\text{STD}=6.987$). We therefore estimate the cost of using \systemname as $\$.021$/image with the ability of running up to $3553$ images/minute using ChatGPT-4o-2024-08-06 as of March 2025\cite{HelloGPT4o}. Researchers note that the time of day can greatly affect the scanning time and reliability as running \systemname during North American working hours would be drastically slower and lead to more dropped API requests. \footnote{Pricing and timing estimates can vary greatly depending on selected LLM. Other LLMs can be used to reduce cost.}

\begin{figure*}[t]
    \centering
    \includegraphics[width=1\linewidth]{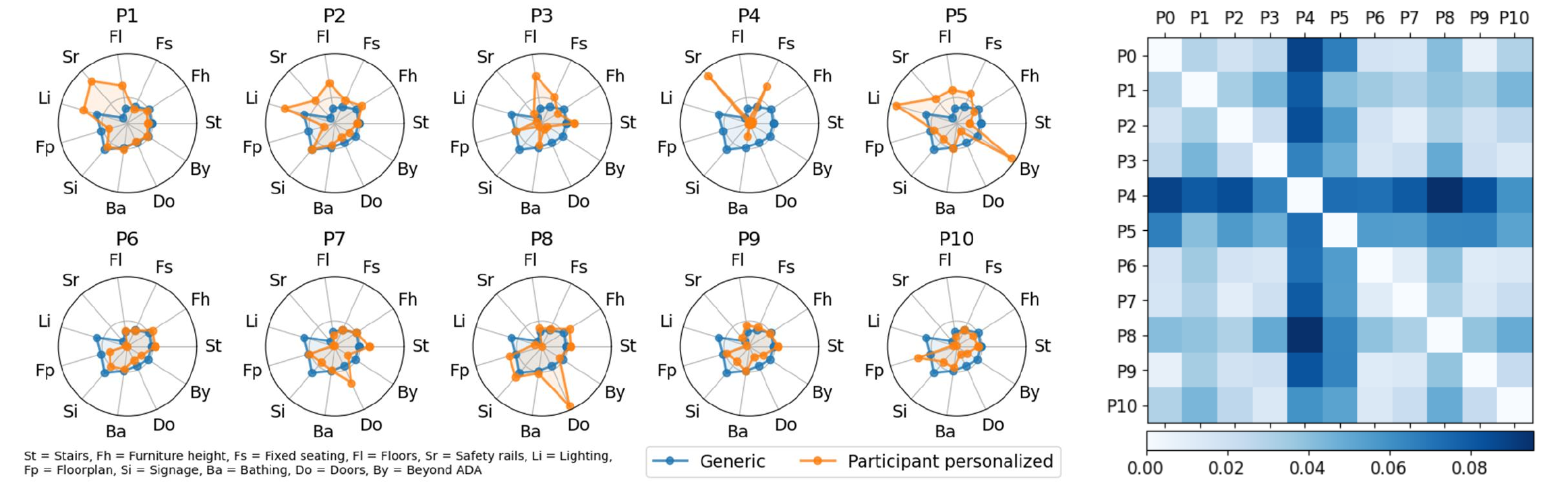}
    \caption{Left: Distribution of participant's generated concerns grouped by ADA categories. Differences in distribution shape indicate differences in \systemname's generation given a specific participant's user model. All data is displayed on a symlog scale. 
    Right: Wasserstein distance between participant's generated accessibility scans. Higher value indicates greater difference between concerns. P0 indicates a "generic" user model or empty JSON, which was not trained by a real user in our study.}
    \Description{Left: Radar plots show the distribution of accessibility concerns for 10 participants across ADA categories, comparing generic and personalized models. Differences in shape highlight how personalized models shift concern emphasis. Right: Heatmap of Wasserstein distances between participants' concern distributions. Most participant pairs cluster below 0.06, while a few exceed 0.08. P4 and P5 show much higher Wasserstein distance values across all participants.}
    \label{fig:personalization-graphs}
\end{figure*}

\subsection{Distribution of Generated Concerns}
To understand the distribution of these generated concerns, we conducted an unsupervised topic clustering of generated concerns using BERTopic \cite{grootendorst2022bertopic}. The top 25 largest clusters were included for further categorization. First, researchers identified a set of key accessibility clusters related to ADA guidelines. Researchers manually assigned these clusters to ADA categories (\eg, Furniture Height, Floorplan), and assigned the rest of the clusters to "Beyond ADA". This process resulted in 11 categories shown in \Cref{fig:concern-diversity}. We find that concerns tend to correlate with types of environments. For example, "Fixed seating" consists of a higher percentage of restaurant concerns ($22.70\%$) and "High, Reach, Difficult" concerns are highly prevalent in grocery stores ($27.31\%$) due to high shelves of items. We found two clusters that went beyond existing ADA categories: "tv, television, remote" and "noise, noise sensitivity, sensitivity". This result indicates the potential for \systemname to extend beyond ADA classifications of accessibility to individual needs of the user. Furthermore, the identification of a "noise" related cluster indicates that \systemname considers \textit{people} surrounding a user with limited mobility as an important yet often overlooked facet, and how environments might change even if "noise" is not directly depicted. Our findings indicate that \systemname aligns with common-sense expectations of where accessibility concerns typically arise, demonstrating how our approach can generate high-quality accessibility data. Moreover, \systemname goes beyond detecting only visible environmental features, enabling dynamic and nuanced spatial inferences.

\label{sec:tech-eval-personalization}
\subsection{Accessibility Scan Personalization}
To evaluate \systemname's personalization, we use the same clustering procedure as the previous section with a new analysis to compute the distribution of accessibility scan clusters across participants (\Cref{fig:personalization-graphs} left). Additionally, we measure the \textit{Wasserstein Distance} between each participant's distribution scaled by the total number of concerns in each category which can be interpreted as the amount of work to transform one participant's concern distribution to another (\Cref{fig:personalization-graphs} right). We note that this metric does not reflect the quality of personalization, which is evaluated in the user study, but rather highlights the variation in accessibility predictions across participants. High Wasserstein distances align with differences in mobility across participants. Notably, P4 and P5 demonstrated the highest average Wasserstein distance ($0.0708$ and $0.0523$) as the only participant who did not have a diagnosed disability and the only participant who walks with the assistance of a walker, respectively. Even among participants with low Wasserstein averages, we still note that there are noticeable differences in their distribution of concerns as shown in \Cref{fig:personalization-graphs} left. 
\revised{For instance, P6, P9, and P10 had a lower average Wasserstein distance ($0.0262$, $0.0286$, and $0.0304$ respectively) as P6 and P9 were both highly independent and believed that they could handle most challenges that came their way, often marking newly identified concerns as irrelevant during training, and P10 who believed that most concerns were not relevant as they were not able to access that environment feature in the first place given their highly limited mobility (\eg, fixed seating was not relevant since they could not transfer at all).}
These results indicate that \systemname can differentiate between the unique needs of individual users to generate meaningfully different accessibility scans.

\label{sec:user-study}
\section{User Study}
We conducted a final user study to better understand the capabilities of \systemname for personalized accessibility scans and the usability of the system by comparing the usefulness between concerns generated from a generic and personalized model. \revised{We believe a generic user model is analogous to static one-size-fits-all approaches to accessibility assessments like checklists while equally susceptible to hallucinations as the personalized model which allows us to evaluate the impacts of personalization on usefulness.} We recruited 10 participants (P1-P10) with varying levels of self-described mobility. Participant demographic information is listed in \Cref{tab:user-study-demographics}. Participants received \$100 compensation for completing the full study, which comprised two stages of approximately one hour each. \revised{Our study was approved by our institution's IRB.}

\subsection{Procedure}
All participants were sent an initial survey requesting basic demographic information and a self-description of their physical and mental capabilities. Their self-description was then used to generate an initial user model. 
A dataset of images of different environments was then compiled through the following procedure. \revised{Participants were first asked to provide 15 different locations they have physically explored to help participants draw from previous experiences and better evaluate the performance of \systemname. For all evaluations, users were asked to take into account any prior knowledge they had visiting the depicted location, implicitly evaluating accuracy as well. Researchers then randomly selected 15 unfamiliar locations. For each location, researchers sourced one image following the same criteria as the technical evaluation data in \Cref{subsec:tech-data-setup}.} Two researchers then conducted the two-stage virtual user study with each participant through online meetings. 

Each study began with an initial one hour stage where participants were asked to train \systemname through a user-guided accessibility scan process. Participants were first given an introduction and briefing on how to use \systemname to evaluate generated concerns and provide their feedback. Participants were given the option of manually controlling the system through remote desktop control or dictating actions to the researchers to use the system. A random selection of 15 images from the compiled dataset was used in this stage. 

Following the first stage of the user study where user's created their own user model, researchers conducted a blind test to analyze how well \systemname personalized to the user. Researchers scanned the remaining 15 images twice using \systemname, once with the trained user model and once with an empty JSON (generic) user model only representing the LLM's innate knowledge of accessibility. Participants were not informed of how the concern was generated (\ie, from the personalized or generic model) until after the study concluded, ensuring a blind test process to eliminate potential bias.

\begin{figure}[t]
    \centering
    \includegraphics[width=1\linewidth]{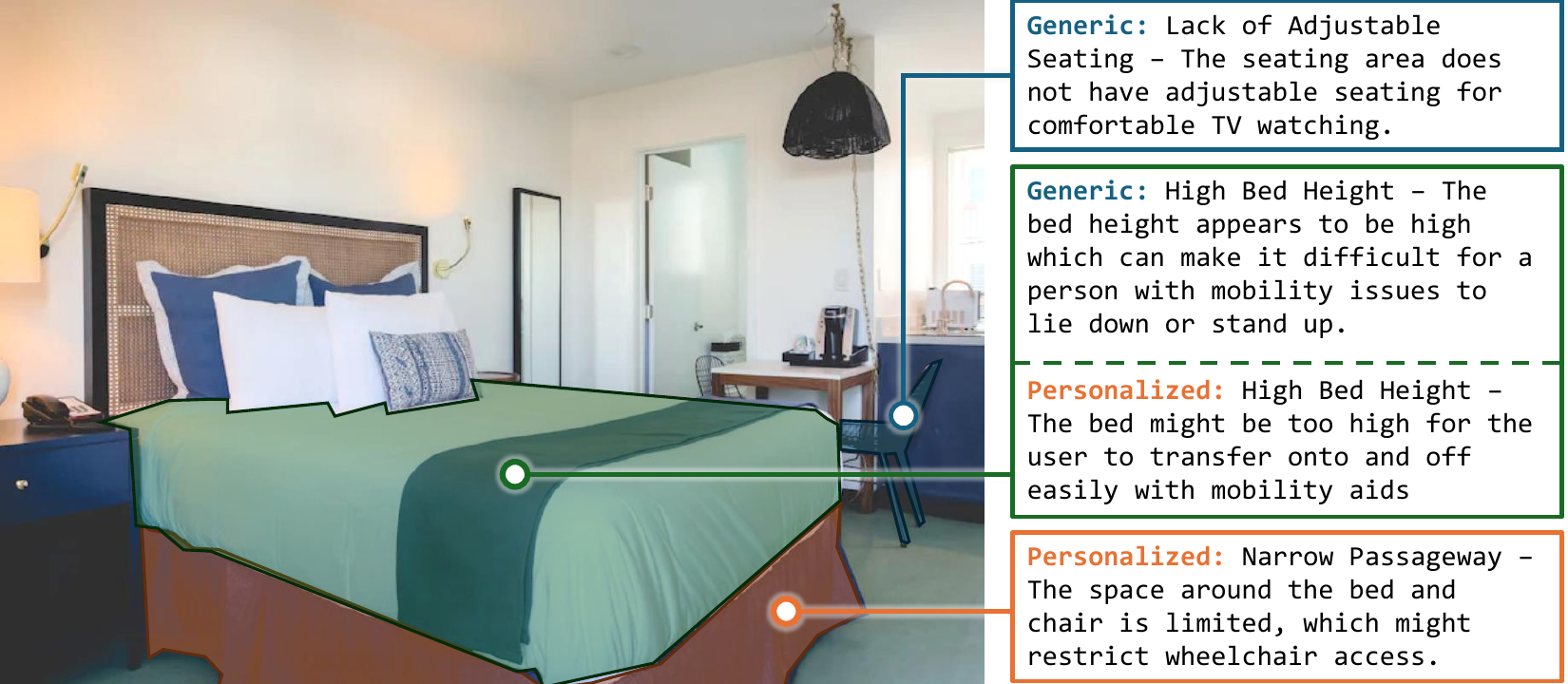}
    \caption{An example of the user study setup for P2. Users are asked to rate the usefulness of \textit{Unique Concerns} (blue and orange) and \textit{Similar Concerns} (green) against each other in a blind test.}
    \Description{Image of a bedroom setup used in a user study for participant P2, showing how concerns from the generic and personalized user model are split in the user study. One generic (lack of adjustable seating) and one personalized (narrow passageway) is shown. A third annotation shows a concern shared by both (high bed height) with differing explanations of why the concern has been identified.}
    \label{fig:user-study-diagram}
\end{figure}

First, 15 \textit{Unique Concerns} (concerns present from one model's prediction but not the other) from each model were randomly selected and shuffled to analyze how well \systemname could identify personalized accessibility needs. An example of different unique concerns is shown in orange and blue in \Cref{fig:user-study-diagram}. Participants were then asked to rate the usefulness of knowing each concern before visiting the environment on a 7-point Likert scale. 

In addition, 15 \textit{Similar Concerns} (concerns present in accessibility scans from both user models differing only in wording) were randomly ordered and visualized side by side to evaluate \systemname's personalization of concern descriptions. An example of similar concerns is depicted in green in \Cref{fig:user-study-diagram}. Participants were then asked to  evaluate the usefulness of knowing the concern before visiting the environment for both shown concerns on a 7-point Likert scale and which concern description they preferred.

After completing all evaluations, participants were informed that concerns were generated from \systemname and which concerns came from the generic vs. personalized user models. They were then interviewed to better understand any Likert scale ratings and discrepancies between the two user models. Participants were also  asked to reflect on their overall experience, concerns about AI in accessibility, and the importance of personalization. These interviews were conducted in a semi-structured format.

\revised{
Three researchers then conducted a codebook thematic analysis \cite{braunReflectingReflexiveThematic2019}, a middle ground approach between structured and reflexive methods, using study recordings, transcriptions, and researcher notes. Researchers first converged on a set of a priori themes from prior research and the needfinding study. The first researcher, who also served as the interviewer, then inductively developed the codebook by reviewing all transcripts and notes in relation to the original themes. The three researchers then collaboratively refined the codebook, resolving disagreements and grouping codes into themes. Our thematic codebook is shown in Appendix \Cref{tab:user-themes}.
}

\begin{figure*}
    \centering
    \includegraphics[width=1\linewidth]{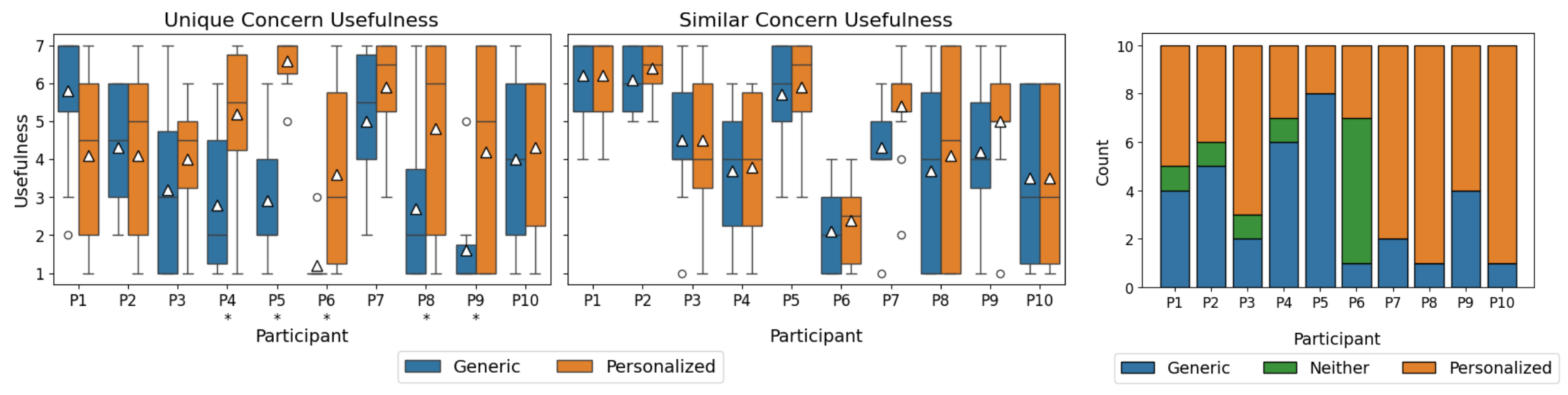}
    \caption{Left: 7-Point Likert Scale scores across the concerns unique to the generic or personalized user model and across the concerns that are present in both the generic and personalized user model. Large white triangles in the box denote means. "1" indicates that the concern was not useful at all and "7" indicates that the concern was very useful. \revised{"*" denotes $P < 0.05$.}. Right: Preferences between generic and personalized models in similar concerns.}
    \Description{Left: Two boxplots of Likert scores (1–7) for unique and similar concerns comparing generic and personalized concerns. Unique personalized concerns generally received higher median and mean ratings across most participants, especially P4, P5, P6, P8, and P9 who had p<.05. Ratings for similar concerns are more balanced but still show higher ratings for personalized concerns. Right: Stacked bar chart showing per-participant preference for "Generic", "Personalized", or "Neither" in the similar concern comparison. Most participants favored personalized concerns with P4 and P5 preferring generic.}
    \label{fig:user-study-graphs}
\end{figure*}

\label{sec:user-study-finding}
\subsection{Findings}
All ten participants were able to train and evaluate \systemname.
\revised{A Wilcoxon Signed Rank Test, a rank-based nonparametric test, was then conducted to assess significant differences between the generic and personalized ratings within each user and across all users. We note that evaluating statistical differences within users can capture more insight into individual user variances at the risk of inflating Type I statistical errors.} \Cref{fig:user-study-graphs} shows participants' ratings and preferences. Across both studies, participants, rating the personalized accessibility scans as relatively useful ($\text{mean}=4.7/7$, $\text{STD}=2.17$).

\revised{Through our thematic analysis, we identified 51 codes (\Cref{tab:user-themes}). Some of the largest codes included ``Accuracy and Detail of Concern'' which highlighted improvements to the accuracy of our system, ``Context Influences Accessibility Concerns'' which captures how users also consider accessibility in regards to when and where something is (\eg, winter vs. summer), and ``AI Usage is Appropriate for Accessibility Scans'' which discusses how users feel comfortable with AI for accessibility assessments}. Below, we present key findings from our thematic analysis. Participant quotes have been lightly edited for concision, grammar, and anonymity.

\textbf{Collaborative AI training is effective.}
As was demonstrated in our technical evaluations (\Cref{sec:tech-eval-personalization}), we found that generated user models from only one hour \systemname's collaborative training was effective in differentiating generated concerns. P2 believed that the current amount of personalization was perfect: \textit{"even with the personalization we just put in, I think is great. I don't think it has to go further than that"} (P2). Furthermore, participants also enjoyed the process of training: \textit{"Oh this is so much fun, I'm loving this"} (P2), \textit{"This is fun"} (P5).
At the same time, participants also noted that the collaborative design of \systemname helped them trust the system more, alleviating concerns on AI accuracy. P8 stated that being able to "double check" the AI was vital: \textit{"when we went through the initial study and we were able to add my input into it, that was amazing like that was absolutely amazing. And you could see the changes from that...I think that's amazing. and I think that's very important"} (P8). P3 shared this sentiment, stating that control was vital for them to use an AI system: \textit{"I'm in control. And in the end I can choose it or not, use it more, use it less, so in the end as long as I have control [over the AI]"} (P3).

\textbf{Personalization generally makes accessibility scans more useful.}
Unique concerns generated from the personalized model were perceived as more useful than those from the generic ($\text{mean}=4.68, \text{STD}=2.26$, $\text{mean}=3.35, \text{STD}=2.24$, respectively (\revised{$p<.001$})), indicating that the addition of personalization generates more useful accessibility scans. P8 echoed these findings, stating that the accessibility scans from the generic user model were laughable: \textit{"[the generic concerns], I would look at that and I would laugh and I would not look at it again you know. But the personalized information that was coming up like I found a lot of that information very useful"} (P8). P10 also found that the addition of personalization made the system more usable by filtering out unnecessary information: \textit{"It would just make it easier if it was more personalized. It'd be less information to filter out...it would make it simpler and more efficient to have it personalized"} (P10). 
While personalization was generally perceived as useful, P4 notes that personalization can actually reduce the usability of \systemname when trying to plan for groups: \textit{"[Personalization is] low importance...having more information also allows me to know that if somebody's coming to visit what I'm looking for...It lets me know for more than just myself"} (P4). 

\textbf{Mixed feedback on referencing user capabilities.}
Furthermore, while personalized accessibility scans were generally perceived as useful, the language used to convey this data had more mixed feedback. In comparing descriptions of the same concern, participants rated the personalized and generic model relatively similarly ($\text{mean}=4.72/7, \text{STD}=2.07$, $\text{mean}=4.4, \text{STD}=2.04$ respectively \revised{($p=.187$)}). We also find that participants only slightly preferred the wording of the personalized concern when compared side by side ($\text{mean}=5.6/10, \text{STD}=2.458$) (\Cref{fig:user-study-graphs} right). 
While viewing a concern which they regarded as not specific enough, P8 believed that generic descriptions are less useful: \textit{"Makes [the concern] just a little bit less [useful] when it's so generic"} (P8). P5 noted that mentioning their capabilities helped draw their attention to key accessibility concerns: "To me [the generic and personalized descriptions] look very similar. But the [mentioning of a] walker is just a red light to me, or an alarm bell" (P5). 
When evaluating a concern about staircases causing fatigue, P4 notes that the personalization of \systemname was drawing conclusions for them: \textit{"Having it say could cause fatigue just thinking out loud seems overly narrow and irrelevant...you don't draw the conclusion for me"} (P4). After reading an explanation for how stairs were not accessible since they used a walker, P3 believed that extra language personalizing the accessibility scans obfuscated the important data: \textit{"I don't need to read whole paragraphs about things. I just need [to see it] and it kind of points [concerns] out"} (P3). \revised{These findings suggest that the representation of accessibility concerns is not as important as identifying them in the first place. This is further compounded as participants appreciated the ability to verify identify concerns, rendering the concern descriptions redundant in many cases.}

\textbf{Potential applications.}
During our user study, participants offered a variety of different use cases for \systemname. P7 viewed \systemname  like a pre-game report to lower the uncertainty in new situations: \textit{"The more I can get advanced scout reports [on accessibility information], the more I can avoid all the uncertainty and angst of the first couple of visits"} (P7). P9 shared similar sentiments, stating that \systemname could help them plan: \textit{"If I know more about the location then I can bring things with me to help overcome the deficits that I do have"} (P9). While P6 believed the system was not as useful for scouting unavoidable environments, they found utility in using \systemname to find places to go to.

\textbf{System improvements.}
During our user study, participants shared detailed suggestions for improving \systemname. All participants believed that the explanation for why a concern was found could be more accurate to their individual needs and the environment. P1, P3, and P7 suggested that the concern descriptions could be made more concise. P8, P9, and P10 noted that the visual highlights were sometimes inaccurate and distracted the user. P3 and P4 believed that the ability for users to see human feedback on the environment would help them trust the system more. P8 and P10 also noted that \systemname would sometimes show irrelevant concerns for tasks they could not complete at all (\eg \systemname showed inaccessible seating when they are unable to transfer seating at all). Finally, researchers also noted that concerns would occasionally duplicate, signaling a need to improve concern concatentation in future versions of our system.

\section{Discussion}
We discuss key implications of our findings, limitations, and opportunities for future work.


\textbf{Application scenarios.}
\systemname's LLM-based approach supports highly personalized and scalable accessibility auditing, appealing for a variety of different use-cases. Through our formative and user study, we identify some potential use cases.

\textit{Prior-visit auditing.}
Uncertainty about a space's accessibility often deterred participants from engaging in activities and made potential visits more daunting and stressful. \systemname is a practical solution to this problem, allowing users to preview potential accessibility concerns before their visits. Users could use \systemname to plan what assistive devices to take, decide whether to travel with a partner or caretaker, or guide deeper inquiries into specific accessibility risks.

\textit{Prior-visit location selection.}
Difficulty finding accessible dining or lodging often discouraged participants from trying new places. \systemname simplifies this process by allowing users to run scans across on publicly available images of environments to generate easily skimmable accessibility insights that can guide future trips and uncover new places to visit. \systemname's scalability can also enable future work conducting large-scale analysis of the accessibility of built environments across regions.

\textit{Sharing lived experiences.}
Participants were excited that \systemname could help them share parts of their lived experiences that are often hard to describe, building empathy and understanding among those around them. In doing so, participants believed they could better prepare the people around them to be more informed accessibility auditors. For other people experiencing major life changes like illness or injury or their loved ones, \systemname can reduce the uncertainty of adapting by providing new perspectives on what challenges in built environments they may encounter in the future. 

\textit{Improving spaces for new demographics.}
While existing accessibility tools like ADA checklists \cite{ADAChecklist2024} or RASSAR \cite{su2024rassar} try to capture as many people as possible in its static definitions of accessibility, \systemname allows building owners and businesses to identify key accessibility concerns for specific demographics of people by evaluating environments on a specified user model, enabling a more targeted approach for space design. For instance, Airbnb owners can use \systemname to rearrange furniture for the specific needs of their guest or government officials can use \systemname to evaluate federal housing for target groups.

\textbf{Perspectives on personalization.}
Our user study found that personalized data was perceived as more useful (\Cref{fig:user-study-graphs} left). With only a brief one-hour personalization session, \systemname generated user models appreciably different from baseline (\Cref{fig:personalization-graphs} left). We also note that participants had varying preferences for the way accessibility concerns are described. Given this, we believe that further personalization is necessary to adapt not only the data, but its representation to the preferences of the user. \revised{These results support the findings of previous works in perceived accessibility \cite{potPerceivedAccessibilityWhat2021, curlSameQuestionDifferent2015, lattmanNewApproachAccessibility2018, mccormackObjectivePerceivedWalking2008, vandervlugtWhatPeopleDeveloping2019}, which state that the willingness to travel to somewhere is highly dictated by an individual's perception of that environment's accessibility.} Personalization offers a new approach to accessibility assessment by capturing a small slice of how they ``see'' to better measure whether or not they would really want to go. Beyond usability, participants also stated that personalization made them feel more heard. As P2 states, \textit{"That's so cool because the user right away feels like they have a voice and they're being heard like hey this is a concern for me. So thats super cool"} (P2). Thus, we believe that personalization can be an important tool in building adoption for future AI accessibility systems by validating user experiences. While we take one approach using structured JSON to personalize LLM systems, future works should explore other representations and methods for user modeling like vector databases, retrieval-augmented generation, and post-training.

\revised{\textbf{Limitations of using only images. }
\systemname was designed around only environmental images to leverage the wealth of publicly available data from the internet. However, relying solely on images also limits \systemname to only capturing general heuristics in comparison to exact measurements or environment dynamics. Thus, the quality of our predictions is limited by when an image was taken (\eg, a path with snow in the winter vs. in the summer), how reflective the image is of the actual experience of being there (\eg, professionally shot AirBnB pictures vs. user generated content), and what is shown.} While participants in our user study felt that generated accessibility scans were detailed and accurate enough to be perceived as useful, we believe that \systemname serves more as a general-purpose tool to alert users to potential concerns over a comprehensive accessibility auditing system like RASSAR \cite{su2024rassar}. Further research in HCI and AI is needed to quantify and evaluate the quality of accessibility scan given personalization to build better ground truths and guide the development of future AI systems. We envision four future areas of work: 1) How can data accuracy from LLMs be improved through improved post-training, computer vision tooling, and model selection? 2) How can we improve an LLM's ability to make conjectures about non-visual properties from visual cues? 3) \revised{\systemname can be easily extended to include multiple images and other textual data by inputting more data into the context window. What kind of accessibility information can be collected to improve concern predictions other readily available data sources (\eg, user reviews, booking location, time of visit)?} 4) What level of assessment detail is needed for users to evaluate the accessibility of environments?


\revised{\textbf{Explainability in AI systems for accessibility.}
Given that AI technologies are still new and unexplored, many participants in our user study were wary of new AI technologies, especially as many existing accessibility technologies were not applicable to their own needs. By designing \systemname to mimic how users think about accessibility assessment using a task affordance perspective, users are more able to engage with and understand how each concern was generated. Findings from our user study reflected this, where participants were more receptive to incorrect concerns as they could trace back the reason the concern was generated, felt more engaged in the accessibility assessment process, and generated concerns that they would actually encounter when entering the environment. This design principle closely follows Miller \textit{et al.}'s \cite{millerExplainableAIBeware2017} call to action for new explainable AI systems to avoid the "inmates running the asylum" problem, when systems are designed around researcher needs over the intended user's, and integrate existing models of how people generate, select, present, and evaluate explanations and decisions in AI systems. Prior work has also has demonstrated this, showing that AI trust and transparency are directly related \cite{atfTrustCorrelatedExplainability2025, knowlesManyFacetsTrust2022}. We believe that future LLMs systems, especially those that support the diverse needs of groups like people with disabilities, should continue to be designed around explainable prediction pipelines, which can make them more approachable for new users. In doing so, users can be more seamlessly integrated into future pipelines through human-AI collaborations, increasing the user tolerance for errors and hallucinations which are currently unavoidable in modern LLMs.}

\section{Conclusion}
In this paper, we introduce \systemname, an AI system that offers a new approach to generating personalized accessibility scans at scale. \systemname uses human-AI collaborations to allow users to easily and effectively update their personalization by validating generated assessments. \systemname can take in images of any environment cheaply and quickly, making it uniquely equipped to conduct personalized accessibility scans at a greater scale. Our technical evaluations demonstrate that \systemname can effectively capture a wide range of different accessibility features and adapt to the varying needs of different users. Furthermore, our user studies demonstrate that not only did users find \systemname useful, but the addition of personalization enabled by \systemname improves the overall usability of the data. Through our work, we demonstrate how AI technologies can be used to build scalable personalized accessibility solutions by applying this approach to accessibility auditing, introducing new ways we can build inclusive spaces and technologies alike.

\begin{acks}
This research was supported by the University of Washington's Center for Research and Education on Accessible Technology and Experiences (CREATE) and the UCLA Digital Technology and Solutions OpenAI Project Grant. We give special thanks to our study collaborators and participants for their participation and feedback.
\end{acks}

\sloppy
\bibliographystyle{ACM-Reference-Format}
\bibliography{ref.bib}
\clearpage
\appendix
\newpage
\appendix
\renewcommand{\thefigure}{\thesection.\arabic{figure}}

\renewcommand{\thetable}{\thesection.\arabic{table}}

\setcounter{figure}{0}
\setcounter{table}{0}

\renewcommand{\arraystretch}{1.1}

\section{Formative Study Demographics}

\begin{table}[H]
\caption{Demographic information of six participants (U1-U6) in the formative study. }
\label{tab:needfinding-demographics}
\resizebox{\linewidth}{!}{
\scriptsize
\begin{tabular}{@{}lllp{1.25cm}p{2cm}|llll@{}}
\toprule
 & & & & & \multicolumn{4}{c}{No. Preferred} \\ \cline{6-9}
ID & Age & Gender & Diagnosed Disability & Self-Described Motor Capability & Total & Personalized & Equal & Generic    \\ \midrule
U1 & 41  & M      & Paraplegic                  & Able to use hands fully, paralyzed from the waist down & 4 & 3 & 1 & 0 \\
U2 & 29 & F & Leg amputation                           & Limited to moving around certain terrains & 5 & 2 & 1 & 2 \\
U3 & 53 & M & Spinal cord injury - T11/T12, paraplegic & Over long distances I use my wheelchair, for short distances I use crutches & 4 & 3 & 1 & 0                                \\
U4 & 28  & F      & Multiple sclerosis          & Numbness on my arms and feet, a lot of fatigue & 4 & 2 & 2 & 0  \\
U5 & 47  & M      & C6 incomplete quadriplegic  & No movement from the chest down, C6 and below & 4 & 2 & 1 & 1\\
U6 & 50 & M & Spinal cord injury - C5/C6               & Paralysis below mid chest, limited hand movement, limited wrist flexion, no triceps & 2 & 0 & 2 & 0 \\ \bottomrule
\end{tabular}}
\Description{Table listing demographic details of six formative study participants (U1–U6), including age, gender, diagnosed disability, self-described motor capability, and concern preference counts. Participants have various mobility-related conditions such as paraplegia, leg amputation, spinal cord injuries, and multiple sclerosis. Most participants preferred personalized concerns over generic ones. Personalized concerns were preferred in 2 to 5 cases per participant, with only one instance of a participant preferring generic over personalized.}
\end{table}

\section{Prompts}
\label{sec:appendix-prompts}
\begin{figure}[H]
\centering
\includegraphics{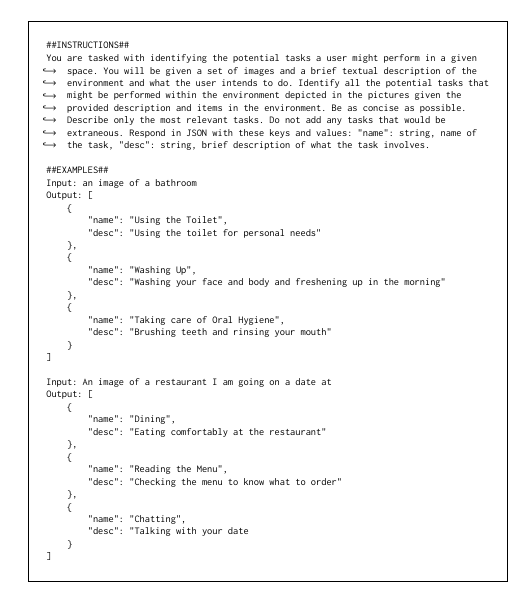}
\caption{Prompt used to identify subtasks a user might do in an environment.}
\Description{ Prompt for identifying user subtasks in a given environment. Instructions ask the model to identify only relevant tasks based on images and user intent, responding in JSON format with keys "name" and "desc". Two examples are shown: (1) a bathroom image, producing tasks like "Using the Toilet" and "Washing Up"; (2) a restaurant image with the intent of going on a date, producing tasks like "Dining", "Reading the Menu", and "Chatting."}
\label{fig:appendix-prompts-identify-tasks}
\end{figure}

\newpage
\begin{figure}[H]
\includegraphics{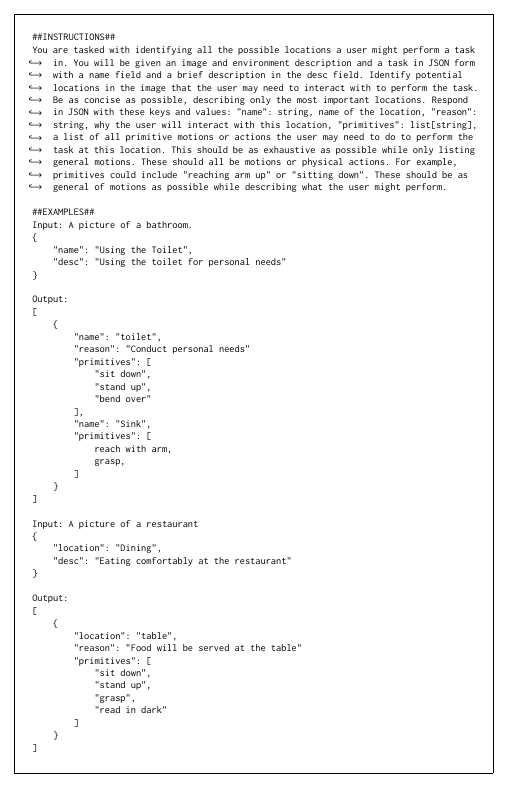}
\caption{Prompt used to identify locations user might perform a subtask and their primitive motions.}
\Description{Prompt for identifying possible locations and associated primitive motions for a user to perform a task in a given environment. Instructions guide the model to output JSON with location name, reason for use, and a list of general motion primitives. Two examples are shown: for a bathroom, locations include "toilet" (with primitives like sit down, stand up) and "sink" (with reach and grasp). For a restaurant, the "table" is identified as the dining location with motions like sit down, grasp, and read in dark.}
\label{fig:appendix-prompts-identify-primitives}
\end{figure}

\newpage
\begin{figure}[H]
\centering
\includegraphics{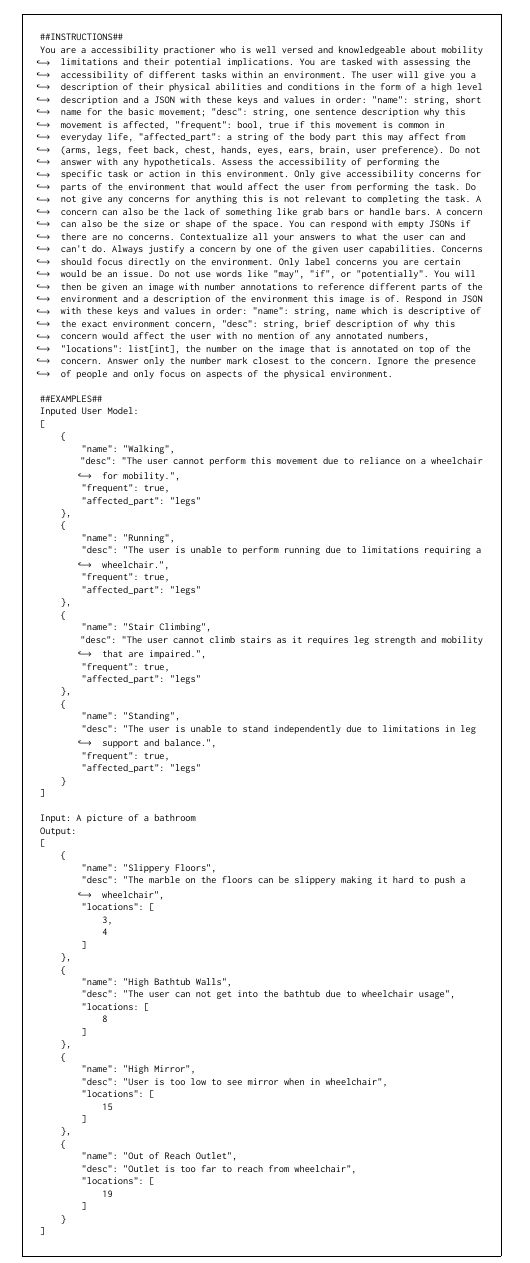}
\caption{Prompt used to identify accessibility concerns in an image.}
\Description{Prompt for identifying accessibility concerns in an image based on a user’s mobility limitations. Instructions describe how to assess concerns using an inputted user model that includes abilities like walking or standing, along with affected body parts. Example user model lists mobility limitations tied to leg function. Given an image of a bathroom, the output identifies concerns such as "Slippery Floors", "High Bathtub Walls", "High Mirror", and "Out of Reach Outlet" each with brief descriptions and location references.}
\label{fig:appendix-prompts-identify-concerns}
\end{figure}

\newpage
\newpage
\clearpage
\section{Thematic Coding}

\begin{table}[H]
\caption{All generated themes and codes from user study.}
\label{tab:user-themes}
\small
\centering
\begin{tabular}{lll}
\hline
Theme & Code & Count \\ \hline
\multirow{2}{*}{User Practices} & Existing Practices For Accessibility Evaluations & 4 \\
 & People Evaluate Accessibility For Their Entire Party & 1 \\ \hline
\multirow{5}{*}{\shortstack[l]{General Concerns\\ About Environment}} & Children Compatible Design & 1 \\
 & Concerns Captured by ADA & 9 \\
 & Concerns Due to Existence of Other People & 4 \\
 & Accessibility Information Availability & 1 \\
 & ADA Is Not Fully Enforced or Maintained & 2 \\ \hline
\multirow{9}{*}{System Usefulness} & Evaluations Have Correlation With Planned Activity & 6 \\
 & Participants Expect Inaccessibilities & 4 \\
 & Context Influences Accessibility Concerns & 10 \\
 & System Identifies Hard to Notice Concerns & 2 \\
 & Can/Cannot Infer Non-Visual Properties from Visual & 6 \\
 & Knowing Accessibility Is More Useful When Choosing a Locations & 1 \\
 & Data Availability Affects Usefulness of System & 6 \\
 & Commonly Accepted Concerns Are Not Useful & 2 \\
 & Build Understanding if Disabilities & 6 \\ \hline
\multirow{6}{*}{System Features} & Able to Accept New Concerns & 1 \\
 & Structured Presentation of Information & 3 \\
 & Training Process Is Positively Perceived & 7 \\
 & Image and Image Highlighting Are Useful & 3 \\
 & Usefulness of Concern Reasoning & 3 \\
 & System Is Fast and Responsive & 2 \\ \hline
\multirow{14}{*}{Improvements Needed} & Overfitting & 1 \\
 & Delay & 1 \\
 & Hallucinations & 2 \\
 & Highlighting Is Inaccurate & 3 \\
 & Duplicated Concerns & 1 \\
 & User Needs Are Dynamic & 4 \\
 & Detail Level of Concern Descriptions & 8 \\
 & Accuracy And Detail of Concern Reasoning & 13 \\
 & Control of What Concerns Are Shown & 2 \\
 & Accessibility Accomodation Recommendations & 1 \\
 & Clickable Highlights & 1 \\
 & Directly Query AI About Places & 1 \\
 & Integrate Other Human Feedback & 2 \\
 & Use Reference Measurements in Concern Descriptions & 1 \\ \hline
\multirow{9}{*}{AI Usage} & AI Usage is Appropriate for Accessibility Scans & 9 \\
 & Concerns on AI Accuracy & 4 \\
 & Concerns on Data Security in AI & 6 \\
 & AI Capabilities of Capturing Perception & 4 \\
 & Fear Of AI Influencing Perception & 3 \\
 & AI Mimicking User Perception Is Useful in Accessibility Scans & 3 \\
 & AI Mimimcking Perception Allows Word View to be Shared & 1 \\
 & Importance Of Mobility Modeling Accuracy in AI & 5 \\
 & Lack Of Trust in AI Prescriptions Of Mobility & 4 \\ \hline
\multirow{6}{*}{Personalization} & Personalization Builds Trust in AI & 2 \\
 & Everyone Is Unique & 4 \\
 & Concerns Should Mention User Capabilities & 9 \\
 & Personalization Make People Feel Heard & 2 \\
 & Personalization Reminds People of Their Disabilities and Makes Them Uncomfortable & 3 \\
 & Mixed Feelings On Usefulness of Personalization & 8 \\ \hline 
\end{tabular}%
\Description{Table showing thematic coding results from the user study, organized by six major themes: User Practices, General Concerns About Environment, System Usefulness, System Features, Improvements Needed, AI Usage, and Personalization. Each theme includes several codes with counts indicating how often each was mentioned. High-frequency codes include "Context Influences Accessibility Concerns" (10), "Accuracy And Detail Of Concern Reasoning" (13), and "AI Usage Is Appropriate For Accessibility Scans" (9). The table reflects a broad range of user feedback, from concerns about system limitations to trust in AI and the value of personalization.}
\end{table}

\end{document}